\documentclass[11pt,a4paper]{article}
\usepackage[utf8]{inputenc} 
\usepackage[english]{babel}
\usepackage{graphicx,amssymb,amsmath,wasysym,subfigure}
\usepackage{epstopdf}
\usepackage[bookmarks=false]{hyperref}
 \usepackage{mathrsfs}
     {\arraycolsep 0.14em\begin{eqnarray}}{\end{eqnarray}}
\def\be{\begin{equation}}
\def\bea{\begin{equation}}
\def\beq{\begin{eqnarray}}
\def\eeq{\end{eqnarray}}
\def\bea{\begin{equation}}
\def\beqn{\begin{eqnarray}}
\def\ee{\end{equation}}
\def\eea{\end{equation}}
\def\eeqn{\end{eqnarray}}
\def\ba{\begin{array}}       
\def\ea{\end{array}}

\def\lsim{\;\raise0.3ex\hbox{$<$\kern-0.75em\raise-1.1ex\hbox{$\sim$}}\;}
\def\gsim{\;\raise0.3ex\hbox{$>$\kern-0.75em\raise-1.1ex\hbox{$\sim$}}\;}
\def\beq{\begin{equation}}   \def\eeq{\end{equation}}
\def\ba{\begin{array}}       \def\ea{\end{array}}
\def\bea{\begin{eqnarray}}   \def\eea{\end{eqnarray}}
\def\nn{\nonumber}
\def\k{\kappa}
\def\l{\lambda}

\def\issue(#1,#2,#3){{\bf #1}, #2 (#3)}
\def\PREP(#1,#2,#3){Phys.\ Rep. \issue(#1,#2,#3)}

\newcommand{\M}{\mathcal{M}}

\def\mygraph#1#2{ \subfigure[]{
   \label{#1}
   \hspace*{-0.5in}
   \begin{minipage}[b]{0.45\textwidth}
   \centering
   \hspace*{4ex}
   \includegraphics[width=1.10\textwidth,height=1.10\textwidth]{#2}
   \vspace*{-4ex}
   \end{minipage}}
   \vspace*{-1ex}}
\begin{document}
\begin{flushleft}
\end{flushleft}
\baselineskip=15.5pt

\thispagestyle{empty}


\vspace{.5cm}

\begin{center}

{\Large\sc{\bf 
Dominant production of heavier Higgs bosons through vector boson fusion in 
NMSSM  
}}

\vspace*{9mm}
\setcounter{footnote}{0}
\setcounter{page}{0}
\renewcommand{\thefootnote}{\arabic{footnote}}
\mbox{ {\sc Debottam Das}} \\
\vspace*{0.9cm}

{\em Institute of Physics, Bhubaneswar, Odisha 751005, India \& Homi Bhabha National Institute,
Training School Complex, Anushakti Nagar, Mumbai 400085, India}\\

\vskip5mm
E-mails: {\tt debottam@iopb.res.in}
\vskip5mm

\end{center}

\vspace{1cm}

\begin{abstract}
We study the features of the additional Higgs bosons in the Next-to-Minimal Supersymmetric Standard Model where
the lightest  
beyond Standard Model Higgs boson does not dominantly couple to up-type quarks. The new state
is dominantly singlet-like while it can also accommodate a small
down-type Higgs component.
The gluon-gluon fusion cannot be adequate enough for such a Higgs production.
We show that the
vector-boson fusion may become the leading production mechanism
to probe this new scalar at the LHC. Using the existing
13 TeV LHC data for an integrated luminosity $36.1~fb^{-1}$, we show the LHC constraints on the parameter space. 
Finally, we also study the reach of the planned high 
luminosity LHC (${\mathcal L}=3~{\rm ab}^{-1}$ at $\sqrt s=$~14 TeV) and 
the proposed
high energy upgrade of the LHC (${\mathcal L}=15~{\rm ab}^{-1}$ at $\sqrt s=$~27 TeV) to probe this singlet-like Higgs scalar. 
\end{abstract}

\section{Introduction}
\label{sec:intro}

The Higgs boson of the Standard Model (SM) which is considered as one of the main motivations of the Large Hadron Collider (LHC)
has already been discovered a few years ago \cite{Aad:2015zhl,HiggsDiscoveryJuly2012} with mass $m_H \simeq 125~$GeV. Accommodating the observed Higgs scalar has put significant constraints on 
the allowed parameter space of any existing model of
beyond Standard Model (BSM) physics. In the context of the 
simplest Supersymmetric model (SUSY) – the Minimal Supersymmetric Standard Model 
(MSSM) \cite{SUSYreviews1,SUSYbook1}, large stop masses and/or mixing 
have been required to generate a Higgs mass of $125$
GeV (for a review see \cite{djouadi}) which may lead to dangerous
charge and color breaking minima \cite{ccb}
and could produce a large fine-tuning on the allowed parameter space 
\cite{FT}. The MSSM has an enlarged Higgs sector with CP-even, CP-odd and charged Higgs bosons \cite{djouadi} which has a particle spectrum and couplings as in a generic Two Higgs Doublet Model of type II \cite{Branco} but determined by the MSSM parameters \cite{djouadi}\footnote{Two Higgs doublet extension of the standard model may lead to other interesting phenomenology, see e.g. 
\cite{Zarikas:1995qb,Lahanas:1998wf,Aliferis:2014ofa}}.

In the Next-to-Minimal 
Supersymmetric Standard Model (NMSSM) (for reviews see \cite{review1,review2}), a  
SM singlet superfield $\hat S$ is introduced in addition to the MSSM 
Higgs 
superfields $\hat H_d, \hat H_u$. The solution to the fine-tuning problem can be more easily addressed since the SUSY breaking scale can be made 
relatively low while satisfying the SM-like Higgs boson mass of 125 GeV. 
The $\mu$ term is generated
dynamically by the vacuum expectation value (vev) of the real 
scalar component of the gauge singlet 
superfield $\hat S$. Here one can efficiently enhance the Higgs boson mass either via new tree level contributions \cite{review1,review2,higgs1} or through singlet-doublet mixings 
\cite{higgs2}. Thus large 
$\tilde t$ mass and/or $\tilde t_L-\tilde t_R$ mixings would not be necessary
which can in turn significantly reduce the
fine-tuning \cite{nmssmft}. 
NMSSM offers interesting collider phenomenology as lighter SUSY particles 
are now allowed with a Higgs boson mass of 125 GeV and the fine-tuning 
criterion 
\cite{Cao,nmssmhiggslhc1,nmssmhiggslhc2,nmssmhiggslhc3,nmssmhiggslhc4,nmssmhiggslhc5,nmssmhiggslhc6,nmssmhiggslhc7,Beuria1,nmssmhiggslhc8,Guchait,nmssmhiggslhc9,Beuria2,Guchait2,nmssmhiggslhc10,Baum1,Das1,nmssmhiggslhc11,nmssmhiggslhc12}. Similarly, a singlino-like lightest Supersymmetric particle (LSP) can quantitatively relax the lower bound on SUSY particles through additional SUSY cascades into singlino LSP \cite{singlinonmssm1,singlinonmssm2,singlinonmssm3}. In the scalar sector, 
a very light Higgs is not yet excluded by
the LEP searches \cite{LEP}, provided
its couplings to the SM particles are small enough. 
The extra singlet scalar may lead to 
new decays of Higgs bosons, particularly into two lighter Higgs scalars
 which can qualitatively be different than 
a two Higgs doublet Supersymmetric Standard model \cite{magg1,nmssmhiggs1,nmssmhiggs2} like MSSM.

The
leading-order production processes for the Higgs boson production can be divided as:
(i) $gg \to H$ (ggF), (ii) vector boson fusion (VBF) and associated $V + H$ production 
(iii) associated ${\bar t}t + H$ along with a single $t + H$ production 
(see reviews like \cite{Djouadihiggs,Reina:2005ae,Ellishiggs}). 
The common lore is that the production cross section 
for $m_H \lesssim 1$~TeV is primarily dominated via the gluon-gluon fusion \cite{deFlorian:2016spz} : $gg \to H$ 
through intermediate quark loops, in particular via the top quark. 
For $m_H \lesssim 100$~GeV,
the associated-production
$gg, q {\bar q} \to W + H, Z + H$ can have significant effects
in production. 
The vector-boson fusion processes
$W^+ W^-, ZZ \to H$ become important for $m_H \gtrsim 100$~GeV and dominate the production processes for $m_H \gtrsim 1000$~GeV \cite{vbf}.
Other processes like
$gg, q {\bar q} \to b {\bar b} H$, $t {\bar t} H$
and in association with a single $t$ or ${\bar t}$ could also contribute to 
Higgs productions.

In the case of MSSM, it is known that 
production of the heavier Higgs scalar has also been 
dominated by the ggF for small or moderate values of $\tan\beta$ 
\cite{djouadi,jeremie1,jeremie2}. At high $\tan\beta$, in addition to the ggF 
(where $b$ quark loop now also contributes),
the associated Higgs production with the $b$ quark would also significantly 
contribute, thanks to  
strong enhancement of the Higgs couplings to the down type fermions. 
Consequently, the new Higgs can decay into 
$\tau \tau$ which has been considered as an important
search channel at the LHC. In the case of NMSSM, in the CP-even Higgs sector, in addition to a SM-like Higgs, one  may observe a MSSM-like and also
a new BSM Higgs scalar.  
In a specific parameter space, this extra Higgs scalar 
can be found to be
almost independent of $H_u$ and determined mainly 
by $S$ and $H_d$. Due to a very stringent constraint on the CP-even BSM Higgs couplings to the vector bosons, in the phenomenological acceptable region, the
new state is found to be mostly singlet like.
Then for small $\tan\beta$, we may observe a few new possibilities for such a CP-even BSM Higgs state $H$ with a mass much smaller than the TeV scale ($m_H~<~1$~TeV):
\begin{itemize}
\item 
First we review the role of gluon-gluon fusion in the context of production of
the scalar $H$. It depends on the effective $Hgg$ coupling which in turn 
depends on the masses of the 
squarks and the heavy fermions in the loops.
In the parameter space where squarks are heavier (which has been considered here), the effective $Hgg$ coupling
may not receive significant contributions from SUSY scalars in the loops. 
Similarly, the contributions from the 
heavy fermion loops in the said coupling are also suppressed as the Higgs state
can only have small couplings with the $t\bar t$ and the
$b\bar b$ fermions. While the suppression in the $Ht\bar t$ coupling is related to an insignificant $H_u$ component in $H$, the smallness of the $Hb\bar b$ coupling can be attributed to the small $\tan\beta$ and large singlet component
\cite{review2}\footnote{We note in passing that this is hardly possible  
in the MSSM as the bilinear mixing renders 
the 2 CP-even Higgses as the admixtures of $H_d$ and 
$H_u$. Also satisfying Higgs mass at small $\tan\beta$ would need large SUSY scale 
in MSSM \cite{jeremie1}}. Thus in this specific region, the gluon-gluon 
fusion may not be adequate enough for such a Higgs production. In addition,
the associated Higgs production with the $b$ quark would also be insignificant \footnote{For moderate or relatively larger values of $\tan\beta$, the situation may be a bit different. 
Radiative corrections may 
enhance the non SM-like Higgs couplings to the $b$ quarks which may,
in turn, push both
ggF
and associated Higgs production with the $b$ quark. Similarly, lighter squarks
could also boost the ggF contributions to the Higgs production rate at the LHC. This part of the parameter space would not be considered here.}.
\item
We know that 
vector-boson fusion processes $W^+ W^-, ZZ \to H$ can be
unsuppressed if any of the doublet components in the Higgs is non-negligible. Indeed, this can be true in the NMSSM parameter space where ggF may not be sufficient enough to produce a Higgs boson. 
Here VBF can be observed as the main production mechanism 
even for probing a lighter non SM-like Higgs boson. 
\item
Interesting changes can be observed even in the decays of the new Higgs scalar in the NMSSM.  
$H\rightarrow \tau\tau$ and $H\rightarrow t \bar t$ (when kinematically allowed)
would now be extremely suppressed which are promising search channels for 
MSSM-like heavy Higgs boson. Note that 
the branching ratios (Brs) into 
$\tau\tau$ can be primarily small
if the heavier Higgs state
is dominantly singlet like. 
Final states with 2 gauge bosons and/or 
Higgs$\rightarrow$ Higgs decays  
(when kinematically allowed) may become important to search such a Higgs
scalar
at the LHC.
\end{itemize}

The aim of this work would be to study (i) the status of such a Higgs scalar
in the light of existing 13 TeV LHC data at 36.1 $fb^{-1}$ 
integrated luminosity,
(ii) prospects of probing this new scalar
in the context of planned high luminosity run of the 
LHC (${\mathcal L}=3~{\rm ab}^{-1}$ at $\sqrt s=$~14 TeV) and at the proposed
high energy upgrade of the LHC (${\mathcal L}=15~{\rm ab}^{-1}$ at 
$\sqrt s=$~27 TeV).
We note in passing that there already exists a number of models, 
where neutral Higgs bosons
with suppressed or vanishing couplings to the SM fermions
are present at the weak scale. These are commonly known as fermiophobic Higgs bosons and have many generic features which have been considered 
in the literature \cite{fermio}. They have already been
searched by experimental collaborations \cite{exptfer}.
They can be accommodated in type-I two Higgs doublet models \cite{fer1}.
However, our work considers Supersymmetric 
type-II two Higgs doublet models augmented by a singlet and our conclusion can be generalized to a non Supersymmetric version as well.

The paper is organised as follows. In Sec.\ref{sec:nmssm} we primarily 
discuss NMSSM and its Higgs sectors. In Sec.\ref{sec:results}
we show parameter space where rate of production of a new singlet dominated Higgs scalar via the VBF channel
can be comparable or even dominate over the gluon-gluon fusion process.
Sec.\ref{sec:simulation} describes our predictions at the $\sqrt s=$~14 and 
$\sqrt s=$~27 TeV runs of the
LHC. Finally we conclude in Sec.\ref{sec:con}.

\section{Higgs sector in the Next to minimal Supersymmetric Standard Model}
\label{sec:nmssm}

In the scale invariant NMSSM, superpotential $W$ can be read as 
\cite{review2}:
\beq\label{suppot}
W_\mathrm ~=~ h_u\, \widehat{Q} \cdot \widehat{H}_u\;
\widehat{U}^c_R + h_d\, \widehat{H}_d \cdot \widehat{Q}\;
\widehat{D}^c_R + h_e\, \widehat{H}_d \cdot \widehat{L}\;
\widehat{E}_R^c +\lambda {\hat S}\,\hat{H}_u \cdot \hat{H}_d + \frac{\kappa}{3} 
\hat {S}^3~.
\eeq
In the above, the Yukawa couplings $h_u$, $h_d$, $h_e$ and the superfields
$\widehat{Q}$, $\widehat{U}^c_R$, $\widehat{D}^c_R$, $\widehat{L}$ and
$\widehat{E}_R^c$ should be understood as matrices and vectors in
family space, respectively.
The vacuum expectation value (vev) $s$ of the real
scalar component of $\hat S$ generates an effective $\mu$-term
\beq\label{mu}
\mu_{eff} = \l s\; ,
\eeq
which in turn solves the $\mu$-problem of the MSSM.

In the soft-SUSY breaking sector, apart from the ${\cal L}^{soft}_\mathrm{MSSM}$
which contains soft-SUSY breaking terms of the MSSM except the $B\mu$ term,
there are new trilinear interactions as well as mass terms 
involving the singlet field.
\bea\label{soft}
-{\cal L}^{soft}_\mathrm{NMSSM} &=&
-{\cal L}^{soft}_\mathrm{MSSM} 
+ m_{S}^2 | S |^2
+\lambda A_\lambda\, H_u \cdot H_d\; S + \frac{1}{3} \kappa A_\kappa\,
S^3 + \mathrm{h.c.} \; .
\eea
Compared to the MSSM, the gauge singlet superfield $ \hat S$ augments 
new degrees of freedom to CP-even and CP-odd Higgs
sectors. Hence the spectrum contains
(1) 3 CP-even neutral Higgs bosons $H_i, i = 1, 2, 3$, 
(2) 2 CP-odd neutral Higgs bosons $A_1$ and $A_2$ ,
(3) one charged Higgs boson $H^\pm$.

At the tree level, one may characterise the 
Higgs sector by the six parameters, namely, 
\bea
\l,\ \k, A_\lambda,\ A_\kappa,\ \mu_{eff} \ \mathrm{and} \ tan\beta\equiv \frac{v_u}{v_d}.
\eea 
Then after eliminating
$m_{H_d}^2$, $m_{H_u}^2$ and $m_{S}^2$ 
using the minimization equations of the potential, one can read 
the elements of the $3 \times 3$ CP-even mass matrix 
in the basis $(H_{dR}, H_{uR}, S_R)$  \cite{review2,nmssmhiggslhc9}.
\begin{eqnarray}
\label{mhiggs2}
 \M^2_{S,11} &=&	M_Z^2  \cos^2\beta + \mu_{eff} (A_\lambda+\kappa s)\tan\beta\;  ,\nn\\
 \M^2_{S,12}&=&(\lambda v^2-\frac{M_Z^2}{2})\sin 2\beta-\mu_{eff} (A_\lambda+\kappa s)\; ,\nn\\
 \M^2_{S,13}&=& \lambda v \left(2\mu_{eff} \cos\beta - (A_\lambda+2\kappa s )\sin\beta \right))\; ,\nn\\
 \M^2_{S,22} &=&  M_Z^2 \sin^2\beta +  \mu_{eff} (A_\lambda+\kappa s)\cot\beta 
,\nn\\
 \M^2_{S,23} &=&  \lambda v \left(2\mu_{eff} \sin\beta - (A_\lambda+2\kappa s)\cos\beta\right))\; ,\nn\\
 \M^2_{S,33} &=& \lambda A_\lambda \frac{v^2}{2s}\sin 2\beta  + \kappa s(A_\kappa+4\kappa s)\; .
\end{eqnarray}

In the above, one may use $v^2=v_u^2+v_d^2=M_Z^2/g^2\sim (174\ \mathrm{GeV})^2$ 
($g^2=\frac{g_1^2+g_2^2}{2}$ )
and 
\beq\label{higgsvevs}
H_u^0 = v_u + \frac{H_{uR} + iH_{uI}}{\sqrt{2}} , \quad
H_d^0 = v_d + \frac{H_{dR} + iH_{dI}}{\sqrt{2}} , \quad
S = s + \frac{S_R + iS_I}{\sqrt{2}}\; .
\eeq

\noindent
In the neutral CP-even Higgs sector, one of the lighter Higgs scalar 
$H_1/H_2$ 
can resemble with SM-like Higgs boson whose mass can be expressed as
\beq\label{treehiggs}
M_Z^2\left(\cos^2 2\beta + \frac{\lambda^2}{g^2} \sin^2 2\beta\right)
+rad.corrs.+\delta_{mix},
\eeq
where $\delta_{mix}$ \cite{higgs2} represents singlet-doublet mixing. 
Clearly, for small $\tan\beta$ ($\le 2$), there can be 
an enhancement even at the
tree level via $\sim \lambda^2 \sin^2 2\beta$ term 
(see Eq:\ref{treehiggs}) while for larger $\tan\beta$, $\delta_{mix}$ could play the 
crucial role to obtain SM-like Higgs boson mass $\sim 125~$GeV, provided there is a lighter singlet. In the 
present study, the lightest NMSSM Higgs  
$H_1$ becomes SM-like ($\sim$ $H_{125}$) while the lighter and the heavier 
BSM Higgs bosons
$H_2$ and $H_3$ are  
approximately singlet and MSSM-like respectively. Since the 
squark masses are assumed to be large in this analysis,
the lightest SM-like Higgs boson may receive
handful contributions from the radiative corrections.

Before we present numerical results, here, for completeness, we list the
couplings of the 
lighter BSM Higgs $H_2$ 
with the quarks and the gauge bosons at the tree level \cite{review2}.  
\begin{eqnarray}
H_2 t_L t_R^c & : & -\frac{h_t}{\sqrt{2}} S_{2,2} \nonumber \\
H_2 b_L b_R^c & : & \frac{h_b}{\sqrt{2}} S_{2,1} \nonumber \\
H_2 \tau_L \tau_R^c & : & \frac{h_\tau}{\sqrt{2}} S_{2,1} \nonumber \\
H_2 Z_\mu Z_\nu & : & g_{\mu\nu} \frac{g_1^2 + g_2^2}{\sqrt{2}}
(v_d S_{2,1} + v_u S_{2,2})\nonumber \\
H_2 W^+_\mu W^-_\nu & : & g_{\mu \nu} \frac{g_2^2}{\sqrt{2}} 
(v_d S_{2,1} + v_u S_{2,2}). 
\label{couplings}
\end{eqnarray}
Here $S_{2,i}$ ($i=1,2,3$)
refers to down, up and singlet components. As already stated, up-type 
Higgs component in $H_2$ is considered to be extremely 
tiny i.e., $|S_{2,2}| \sim 0$. Thus $H_2$ couplings
with the up-type quarks are vanishingly small. The same couplings with the 
down-type quarks
are small because of singlet dominance (though $|S_{2,1}| >> |S_{2,2}|$) and also for small values of $\tan\beta$. As explained earlier, the 
couplings of our interest would be $H_2VV$ and its maximum allowance 
where $V$ refers to any gauge boson. Now  
defining reduced couplings $C_{H_iVV}$ as the ratio of Higgs couplings to the vector bosons relative to the corresponding couplings of the SM-like Higgs boson, one gets 
$\Sigma_{i=1,2,3} C^2_{H_iVV}=1$. 
Demanding a SM-like Higgs boson in the spectrum, the said 
condition leads to a small allowance for a
singlet-like or any of the 
BSM Higgs couplings to the vector bosons. For a quantitative estimate of the 
allowance, 
one may find that 
$C_{H_{125}VV} \ge 0.83$ \cite{deBlas:2018tjm} at the 3$\sigma$ CL level which is also quite consistent
with Ref:\cite{HcouplingsLHC}. 
The future prospects for the measurements of 
$C_{H_{125}VV}$
at the LHC depend upon uncertainty scenarios in particular and have been discussed in detail 
in Ref:\cite{futureATLAS,futureCMS,Cepeda:2019klc}.
Using Ref:\cite{Cepeda:2019klc}, one may set a lower bound on $C_{H_{125}VV} \sim 0.95$ 
at the 3$\sigma$ CL level which means that $C_{H_{BSM}VV}$ can accommodate an allowance of $\sim 0.3$ (where $H_{BSM}$ refers to any BSM Higgs $H_2$ or $H_3$). 
The said limit has been obeyed in our analysis. This, in general, 
makes VBF processes as somewhat less
interesting to probe the properties of any BSM Higgs scalar at the LHC.

\section{Choice of parameters and LHC constraints}
\label{sec:results}
\noindent
We use the code NMSSMTools \cite{ntools1} to compute masses and the 
couplings for 
sparticles and Higgses and also branching ratios (BRs) of the Higgs states. In the 
present context we focus on the small $\tan\beta$ ($= 2$). 
This choice effectively minimizes the $b$-quark loop contributions to
Higgs production via ggF processes. For different parameters and 
soft SUSY breaking terms we make the following choices:
\begin{itemize}
\item We set squark masses $\sim$ 2.5~TeV in the present 
analysis. This choice is motivated by the fact that any
SUSY particle has not been yet observed at the LHC. With this choice, the 
contributions coming from the squarks mediated loop diagrams in the effective 
$Hgg$ couplings can be neglected \footnote{For smaller
values of squark masses, squarks loops may  
enhance the effective $Hgg$ couplings
which may, in turn, enhance the production rate of a Higgs boson through ggF process.}.  
\item  Trilinear soft SUSY breaking terms are set at $A_t = A_b = A_\tau= -1.6$~TeV.
\item Masses for all gauginos are set at 2~TeV. Here Higgsino can be the lighter
   which can in
general make Higgsino as the viable dark matter (DM) candidate of the Universe. 
But in the case of singlino-Higgsino or Higgsino like DM,
Direct-detection experiments like LUX \cite{lux} can lead to 
stringent constraints on the 
parameter space 
which is particularly severe for a 
lighter Higgs spectra  
\footnote{ Note that 
satisfying the limit
on Spin-independent direct detection cross-section would require some unnatural fine tuning among input parameters. In addition, one has 
to make use of uncertainties that can arise in the calculation of the 
neutralino-nucleon elastic scattering cross-section,
(see e.g.,\cite{astrouncer,myastro}).}. Assuming gravitino as the LSP would relax this tension completely. Note that thermally produced
gravitinos may form the observed dark matter \cite{nmssmgravitino} without conflicting with  
the particle dark matter search.  

\item We impose a requirement that Landau pole singularities of the
running Yukawa couplings would 
not occur below $10$ TeV \cite{review2},  
which may otherwise lead to conflict
with precision electroweak tests of the SM. 
This typically yields a requirement on the coupling parameter $\l$ 
(one needs to satisfy $\l \le 2$). The NMSSM with such a large $\l$ (precisely $0.7 <\l< 2$) is
commonly called the $\l-$SUSY model 
\cite{lsusy1,lsusy2,lsusy3,lsusy4}. In this case, larger 
tree level contributions helps one to accommodate 
a SM-like Higgs boson with mass of 125 GeV more easily. One may also find that the sensitivity of the weak scale to the stop mass
scale would be reduced here \cite{lsusy1,lsusy2,lsusy3}. In this analysis 
we scan $\l \in 0-2$ and $\kappa \in 0-1$ and  
always check the
absence of
unphysical global minima of the Higgs potential as done in NMSSMTools 
\cite{review2}.
\item
We use the following limit for SUSY Higgs mass $m_{H_1}$ which is the lightest  
Higgs particle of the NMSSM spectrum with standard model like couplings.
\begin{align}
\label{Higgs_bound}
122.1 ~ {\rm GeV} &\leqslant m_{H_1} \leqslant 128.1 ~{\rm GeV}. 
\end{align}
A 3~GeV theoretical uncertainty around $m_{H_1} \simeq 125~$GeV has been 
considered due to uncertainties in
the computation of loop corrections up to three loops, top quark mass, renormalization
scheme and scale dependence etc.\cite{loopcorrection}.

\item
On the flavor physics side, the 
constraints from B-physics namely $B \rightarrow X_s +\gamma$, 
$B_s \rightarrow \mu^+ \mu^-$  
have been implemented, as done in NMSSMTools.

\item Finally we use 173.1~GeV for the top quark pole mass.
\end{itemize}

\begin{figure}[!htbp]
\vspace*{-0.05in}
\mygraph{figlk}{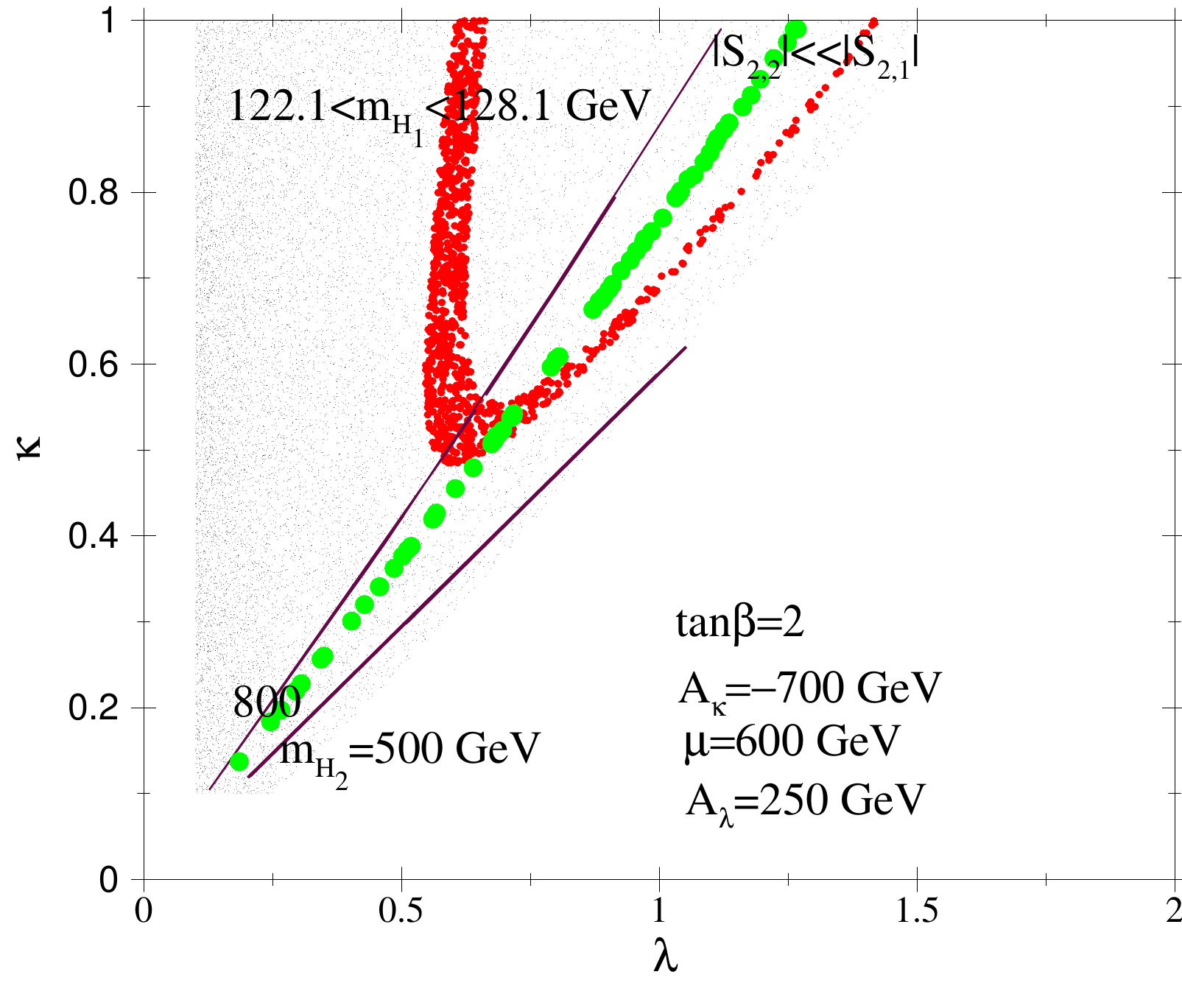}
\hspace*{0.7in}
\mygraph{figkak}{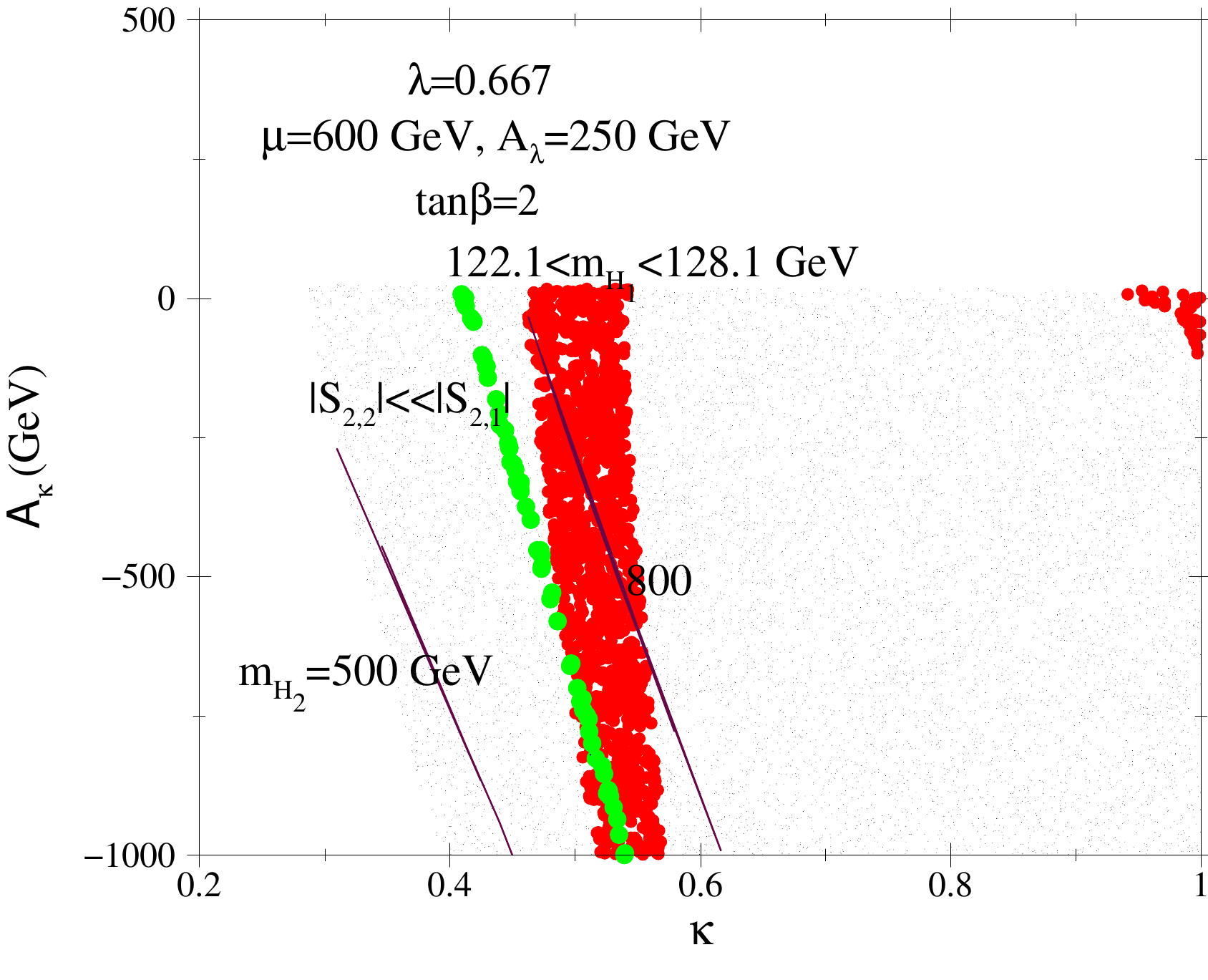}
\caption{(a)Parameter space consistent with the 
Higgs mass and other phenomenological constraints (as implemented in NMSSMTools) has been shown
by the red points in the $\l-\kappa$ plane for 
tan$\beta=2,~A_\kappa=-700$~GeV, $A_\l=250$~GeV and $\mu_{eff}=600$~GeV. 
Gray points are ruled out by one or more phenomenological constraints.  
The green region represents the parameter space in general 
where the absolute value of the 
up-type Higgs component in the heavier Higgs state 
$H_2$ is $\le 0.001$. Along the strip one may obtain 
$\frac{C^2_{H_2VV}}{C^2_{H_2gg}}$ as large as 1000.
(b) Same as fig.(a), but now the parameter space has been studied in the 
$\kappa-A_\kappa$ plane for a fixed $\l=0.667$. 
}
\label{fig1}
\end{figure}

We would now try to see the feasibility 
of the NMSSM parameter space where the ggF can be observed as a sub-dominant
process for the lightest non SM-like Higgs boson production at the LHC. Our aim would be to find 
the parameter space consistent with the experimental and the theoretical constraints and where $H_2$ couplings with the
up-type quarks are vanishingly small. 
We first show our results 
in the $\l-\k$ plane (see Fig.\ref{fig1}a)  
where the input parameters are kept fixed at
tan$\beta=2,~A_\kappa=-700$~GeV, $A_\l=250$~GeV and $\mu_{eff}=600$~GeV. While varying the input parameters, constraints coming from different phenomenological 
observables 
(like one of the 
Higgs boson would be SM-like, collider and the flavor physics constraints), as implemented in NMSSMTools have been checked. 
The gray points in the 
parameter space in the Fig.\ref{fig1}a are in general
not compatible by the said constraints, thus fail to qualify as the valid points. The red region has been drawn to represent the allowed spectra where we 
mainly neglect the  
landau pole or the dark matter constraint. This region satisfies the aforesaid flavor physics
constraints and
is consistent with a SM-like Higgs boson mass with $122.1<m_{H_1}<128.1$~GeV. 
In the figure, the green region represents the parameter space where one may obtain $|S_{2,2}| << |S_{2,1}|$ for the lightest non SM-like Higgs boson $H_2$. 
This region, in general, represents the parameter space where $H_2$ couplings
with the up-type fermions are tiny, in particular, 
one finds that $|S_{2,2}| \le 0.001$. Clearly, the green region mostly falls in the gray part while a small part is consistent with the phenomenological constraints (this part resides on the top of red regions). 
Such a small value of $|S_{2,2}|$ depends on the input parameters that specify the Higgs masses and mixings (see Sec:\ref{sec:nmssm}). From the figure, one may see that the green strip where one can get a tiny $|S_{2,2}|$ corresponds
to $\l \sim \kappa$. Though the exact dependence can only be understood numerically, intuitively the linear dependence in the $\l-\kappa$ plane
can be found
if one sets $\M^2_{S,32}$ $\simeq 0$
in the tree level mass matrix for the CP-even 
Higgs scalars in Eq:\ref{mhiggs2}. The exact dependence depends on the other input parameters e.g., $A_\l$, $\tan\beta$ and $\mu_{eff}$. Besides, we also plot contours for $H_2$ 
mass. Now qualitatively, a tiny $|S_{2,2}|$ refers to suppression in the reduced $H_2gg$ coupling, noted by $C^2_{H_2gg}$. As already discussed, this is the specific region of our interest where the VBF mode may turn out to be the main production 
channel.
For an estimate of the relative dominance of the VBF mode over the ggF channel
in the computation of the production rate of $H_2$ scalar 
we also present the ratio of the their reduced couplings respectively, i.e., 
$\frac{C^2_{H_2VV}}{C^2_{H_2gg}}$. Notably, one may also view the ratio as 
$\frac{\sigma^R(H^{VBF}_2)}{\sigma^R(H^{ggF}_2)}$ where
${\sigma^R(H^{VBF}_2)}$
and ${\sigma^R(H^{ggF}_2)}$ are the reduced production cross-sections in the VBF and the ggF mode respectively 
\footnote{The absolute
value of the cross-section may be obtained if we multiply it with the respective production cross-section of the $H_2$ scalar assuming it SM-like. Typically, for a heavy Higgs having SM-like properties with mass $m_{H_2} \sim O(100)$ GeV, the ggF dominates the production process over the VBF by an order of magnitude i.e., $O(10)$ \cite{xsection}.  
Thus a large value of $\frac{\sigma^R(H^{VBF}_2)}{\sigma^R(H^{ggF}_2)}$ or 
$\frac{C^2_{H_2VV}}{C^2_{H_2gg}}$ 
may clearly define the VBF dominance over the ggF in producing the $H_2$ scalar 
in the NMSSM.}. We find that the ratio becomes maximum 
along the green strip and can be as large as 
$\sim 1000$. Interestingly, this also includes the region which has been
satisfying all phenomenological constraints.
\noindent

Now, it may also be interesting to know the allowance
in $A_\kappa$ to obtain
$|S_{2,2}| << |S_{2,1}|$ or a tiny $|S_{2,2}|$ for a given $\kappa$ 
(see Fig.\ref{fig1}b). To understand the same, we set the
input parameters at 
$\tan\beta=2, \mu_{eff}=600~\rm GeV, A_\l=250~\rm GeV$ which are the 
same as in Fig.\ref{fig1}a. With this choice of input parameters,
we set $\l \sim 0.66$ and vary $A_\kappa$ 
(-1TeV to 1TeV) to study the parametric dependence in the 
$A_\kappa-\kappa$ plane
for obtaining $|S_{2,2}| << |S_{2,1}|$.
Again, the valid parameter space is shown by the red colored regions where
one may easily accommodate the
SM-like Higgs boson mass constraint. Similarly, green region corresponds to
the parameter space where one gets $|S_{2,2}| << |S_{2,1}|$.
Varying $A_\kappa$ would 
effect the singlet
component of Higgs mass matrix which in turn can influence mass and couplings
of the Higgs bosons.
Indeed observing the green points, one finds that $|A_\kappa|$ may
need to be enhanced with $\kappa$ to satisfy $|S_{2,2}| << |S_{2,1}|$. 
Here one may also find that $|S_{2,2}| \le 0.001$. Similarly, a large 
value of $\frac{C^2_{H_2VV}}{C^2_{H_2gg}}$ 
may be obtained along the green region.
Additionally, we also present the contours of $H_2$
mass for representative values (=500, and 800). In summary, 
a tiny $|S_{2,2}|$ or a large value of $\frac{C^2_{H_2VV}}{C^2_{H_2gg}}$
can be observed in the 
$\l-\kappa$ or in the $A_\kappa-\kappa$ plane (see Fig:\ref{fig1}) which 
may correspond to a $H_2$ scalar with
mass $m_{H_2} \sim 600$ GeV.

\begin{figure}[!htbp]
\vspace*{-0.05in}
\mygraph{figmh2}{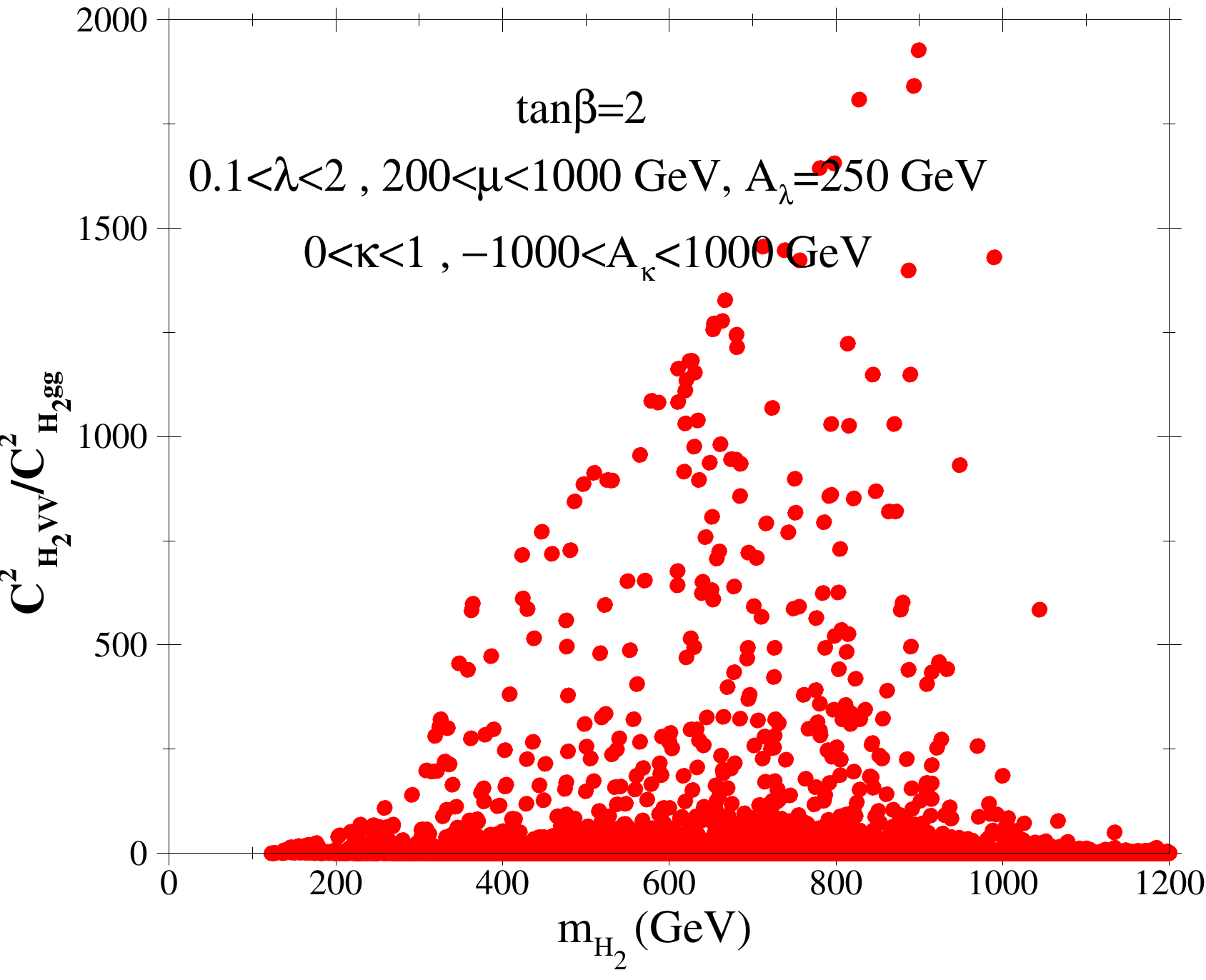}
\hspace*{0.7in}
\mygraph{figcvcg}{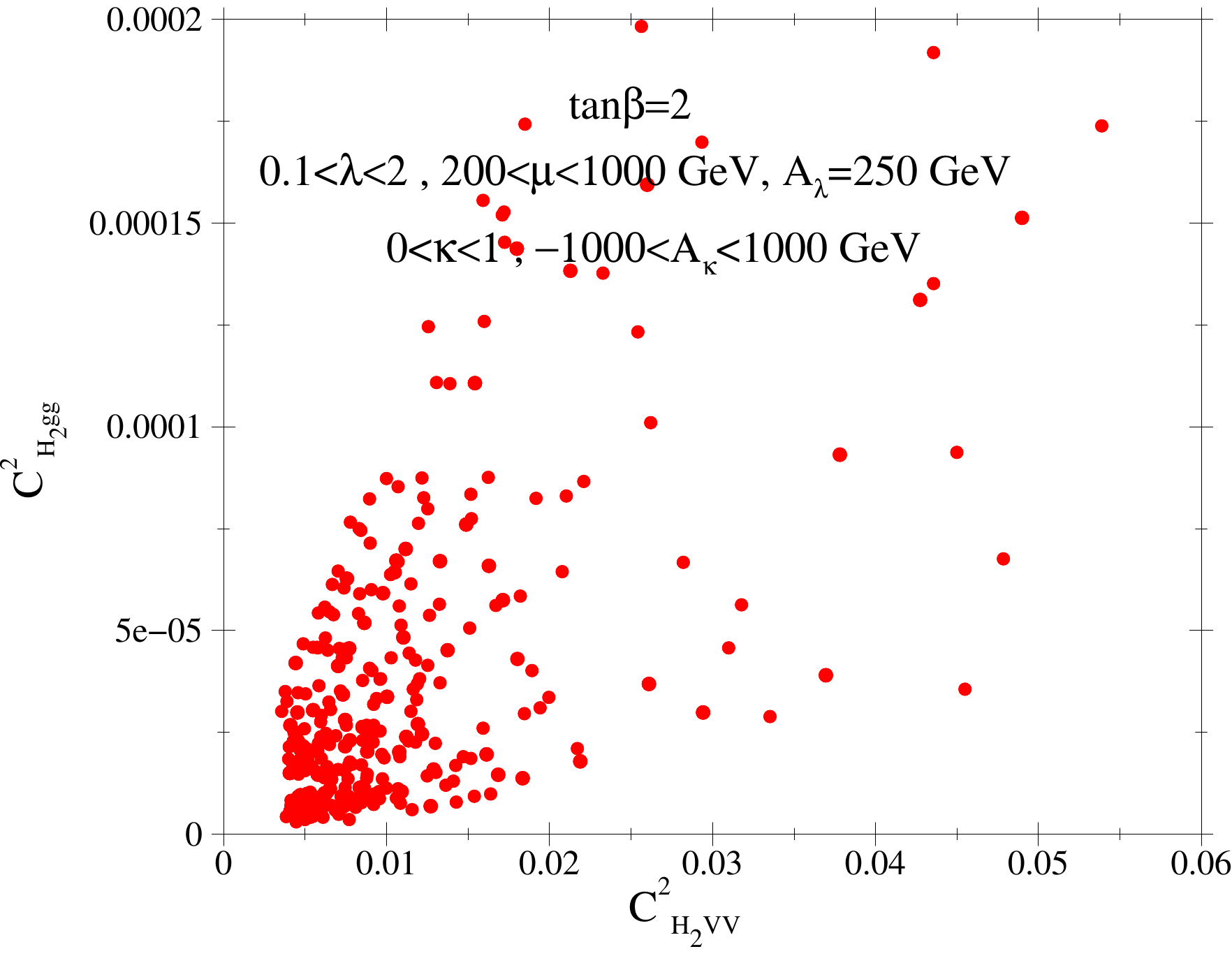}
\caption{(a) $\frac{C^2_{H_2VV}}{C^2_{H_2gg}}$ is shown with mass of the $H_2$ scalar. The ranges of the parameters which we scan are as follows:
$0.1 \le\l\le 2, 0 \le \kappa \le 1, 100 \le\mu_{eff} \le 2000 \rm (GeV), 
-1000 \le A_\kappa\le 1000 \rm (GeV)$. 
The other parameters are tan$\beta=2$, $A_\l=250$~GeV.  
(b) Reduced couplings ${C^2_{H_2VV}}$ and ${C^2_{H_2gg}}$ have been presented
for the same set of input parameters. Here we impose a constraint
$\frac{C^2_{H_2VV}}{C^2_{H_2gg}} \ge 50$.}
\label{figmh2}
\end{figure}

\begin{figure}[ht]
\begin{center}
\includegraphics[width=9cm,height=9cm]{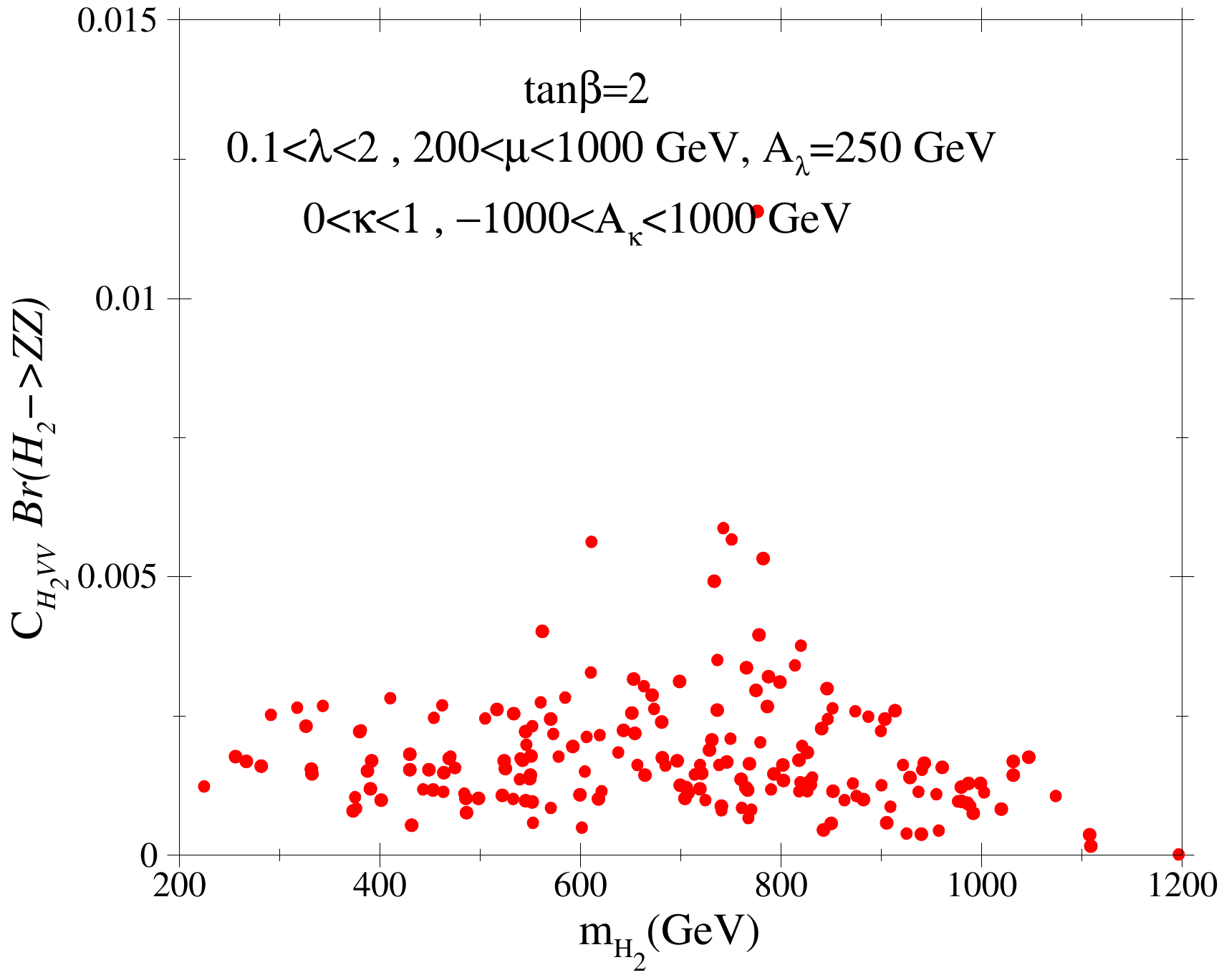}
\caption{Variation of $C^2_{H_2VV}\times Br(H_2\rightarrow ZZ)$ has been shown
  with the Higgs boson mass $m_{H_2}$.}
\label{figcv2mh2}
\end{center}
\end{figure}

We would also analyse if such a large value of the $\frac{C^2_{H_2VV}}{C^2_{H_2gg}}$  
can be accessed over a wide range of $H_2$ mass 
(see Fig.\ref{figmh2}a).  
Here we scan the parameters: 
$0.1 \le\l\le 2, 0 \le \kappa \le 1, 100 \le\ \mu_{eff} \le 2000 \rm ~(GeV), 
-1000 \le A_\kappa\le 1000 ~\rm (GeV)$. 
The other input parameters are tan$\beta=2$,~$A_\l=250$~GeV. 
All the points satisfy the Higgs mass constraint and other phenomenological 
constraints as discussed above. 
The region includes both large and small values of $|S_{2,2}|$. 
In the parameter space where $|S_{2,2}|<<|S_{2,1}|$ 
or where one may obtain a tiny $|S_{2,2}|$ 
a large relative 
enhancement of the $\frac{C^2_{H_2VV}}{C^2_{H_2gg}}$
can be found.
In practice, as discussed already that $|S_{2,2}| \sim 0$ refers to a tiny value of $C^2_{H_2gg}$. Keeping this in mind, 
it may be imperative to estimate the magnitude of the reduced couplings in the 
VBF and ggF mode which we show in Fig:\ref{figmh2}b. 
Here we impose a constraint that
$\frac{C^2_{H_2VV}}{C^2_{H_2gg}} \ge 50$. As can be seen from Fig:\ref{figmh2}b, ${C^2_{H_2VV}}$ may vary up to ${C^2_{H_2VV}}\sim 0.06$.
Finally we also study $C^2_{H_2VV}\times Br(H_2\rightarrow ZZ)$ for the same
set of input parameters
with the mass of the Higgs of our interest in Fig:\ref{figcv2mh2}. 
This may give an idea whether $H_2$ scalar of a given mass 
can be probed 
at the LHC.

In order to demonstrate the observable effects where VBF contribution 
to $H_2$ production can be promising, 
which can potentially be seen at the high luminosity 
run or at the upgraded LHC, we present a few benchmark points (BPs) 
in Table:\ref{tab1}. All the points, shown here, are consistent with constraints
related to the observables in the Higgs sectors as done in 
HiggsBounds-4.3.1 \cite{higgsbound}. We set $\tan\beta=2$ for all these points. In addition, for each BP we show masses for Higgs scalars, 
compositions of scalars (only for the SM-like and the lighter BSM Higgs), reduced couplings 
of $H_2$ with the electro-weak gauge 
bosons,  with the gluons and  with the fermions. In all cases,
$H_2$ is dominantly singlet like 
with vanishingly small $H_u$ components. 
Thus, without any surprise, the 
reduced 
coupling of $H_2$ scalar to the top quark
$C^2_{H_2t \bar t}$ is extremely suppressed. Similarly, the reduced couplings
with the down type fermions are smaller for $\tan\beta=2$. As a natural consequence, one would
expect larger $C^2_{H_2WW/H_2ZZ}$ in comparison to $C^2_{H_2gg}$ which may,
in turn, enhance the relative contribution of the VBF processes in  
$H_2$ production compared to the ggF channel. Indeed one can see it from Table:\ref{table2}, where for the said BPs (as shown in Table:\ref{tab1}),
we present effective production cross-section for the lighter BSM Higgs boson
$H_2$ through the ggF, the VBF and in association with the
$b$ quark channels. Here $\sigma_{SM}$ 
refers to the production cross-section  
corresponding to a Higgs states with mass $m_{H_2}$ having SM characteristics 
\cite{xsection} at the center of mass energy $\sqrt{s}=13~$TeV. For the ggF,
$\sigma_{SM}$ is obtained assuming 
NNLO+NNLL QCD accuracy while for the VBF processes only NNLO QCD 
accuracy is used for the same. Here for the calculation of effective production cross-section we rescale 
the SM cross-sections with the effective squared couplings.

\vskip 0.5cm
\begin{table}[!htbp]
\begin{center}
\begin{tabular}[vt]{clll}
\hline
\hline
parameter & ~~A &~~ B & ~~C  \\
\hline
\hline
$\tan\beta$ &~~2 &~~2  &~~2 \\
$\l$ &~~0.63  &~~0.63 &~~1.35 \\
$\kappa$ &~~0.45  &~~0.46 &~~0.907 \\
$\mu_{eff}$ ~(GeV) &~~267 &~~ 433 &~~  670  \\
$A_{\kappa}$ ~(GeV)&~~-400 &~~ -420 & ~~ -520  \\
$A_{\lambda}$ ~(GeV) &~~250& ~~250& ~~ 250  \\
\hline
\hline
$m_{H_1}$~(GeV) &~~126.3&~~122.5&~~126.2 \\
$m_{H_{2}}$ ~(GeV)&~~250.1&~~500.1&~~700.0 \\
\hline
\hline
$S_{1,1}$ & ~~0.45 &~~0.45  &~~ 0.4 \\
$S_{1,2}$ & ~~0.88 &~~0.89  &~~ 0.9 \\
$S_{1,3}$ & ~~-0.09 &~~-0.07 &~~ -0.2 \\
\hline
\hline
$S_{2,1}$ & ~~0.2 &~~0.18  &~~ 0.34 \\
$S_{2,2}$ & ~~0.004 &~~-0.007  &~~ 0.062 \\
$S_{2,3}$ & ~~0.98 &~~0.98 &~~ 0.93 \\
\hline
\hline
$C^2_{H_2 t \bar t}$ &~~2$\times10^{-5}$ &~~7$\times10^{-5}$  & ~~0.005  \\
$C^2_{H_2 b \bar b}$ &~~ 0.21&~~ 0.17 & ~~0.59  \\
$C^2_{H_2 \tau \tau}$ &~~ 0.21&~~ 0.17 & ~~0.59  \\
$C^2_{H_2 WW/H_2ZZ}$ &~~ 0.0091&~~ 0.006 & ~~0.044  \\
$C^2_{H_2 gg}$ &~~ 0.00013&~~ 0.00009 & ~~0.005   \\
\hline
\hline
$Br({H_2\rightarrow WW})$ &~~ 0.683&~~0.472 & ~~0.363 \\
$Br({H_2\rightarrow ZZ})$ &~~ 0.29&~~0.23 & ~~0.17 \\
$Br({H_2\rightarrow t \bar t})$ &~~ 0.0&~~2.2$\times10^{-3}$ & ~~0.01 \\
$Br({H_2\rightarrow b \bar b})$ &~~ 0.025&~~0.003 & ~~0.0004 \\
$Br({H_2\rightarrow \tau \tau})$ 
&~~3$\times10^{-3}$ &~~4$\times10^{-4}$ &~~6$\times10^{-5}$ \\
\hline
\hline
$Br({H_2\rightarrow H_1 H_1})$ &~~ 0.0&~~0.3 & ~~0.45 \\
\hline
\hline
\end{tabular}
\vskip 0.8cm
\caption{Input parameters for the three benchmark points along 
with masses for the SM-like and the lighter BSM Higgs scalars are shown.
 Different components of $H_2$ scalar
along with the reduced
couplings for Higgs production via the $ggF$, the $VBF$ and associated production 
with the $b$ quark have been presented. 
In addition, we also show the relevant branching ratios.}
\label{tab1}
\end{center}
\end{table}
\begin{table}[!htbp]
\begin{center}\
\begin{tabular}{|c|c|c|c|c|c|c|c|c|}
\hline
\hline
$m_{H_2}$ (GeV)&\multicolumn{2}{c|}{$\sigma(H^{\rm ggF}_2)$~(pb)} 
& \multicolumn{2}{c|}{$\sigma(H^{\rm VBF}_2)$~(pb)}&\multicolumn{2}{c|}
{$\sigma(H^{\rm b\bar bH}_2)$~(pb)}\\

\hline
\hline
& $\sigma_{SM}$ &${\sigma}_{NMSSM}$& $\sigma_{SM}$ &${\sigma}_{NMSSM}$ &
$\sigma_{SM}$&${\sigma}_{NMSSM}$\\
\hline
\hline
BP-A (250) 
& 12.48 & 1.7$\times 10^{-3}$& 1.669 & 1.5$\times 10^{-2}$ 
&4.41$\times 10^{-2}$ & 9.3$\times 10^{-3}$\\
\hline
\hline
BP-B (500) 
& 4.538 & 4$\times 10^{-4}$& 0.4872 & 3$\times 10^{-3}$ 
&2.55$\times 10^{-3}$ & 4.3$\times 10^{-4}$\\
\hline
\hline
BP-C (700) 
& 0.924 & 4.6$\times 10^{-3}$& 0.2275 & 1$\times 10^{-2}$ 
&5.2$\times 10^{-4}$ & 3.1$\times 10^{-4}$\\
\hline
\hline
\end{tabular}
\end{center}
\vskip 0.8cm
\caption{The effective production cross-section for the lighter BSM Higgs boson
$H_2$ for the BPs in Table:\ref{tab1} through the ggF, the VBF and in association with the
$b$ quark have been presented. Here
$\sigma_{SM}$ 
refers to the production cross-section  
corresponding to a Higgs states with mass $m_{H_2}$ having SM characteristics
\cite{xsection} at center of mass energy $\sqrt{s}=13~$TeV. For the ggF,
$\sigma_{SM}$ is obtained assuming 
NNLO+NNLL QCD accuracy while for the VBF processes only NNLO QCD 
accuracy is used for the same.
}
\label{table2}
\end{table}
As $H_2$ is dominated by singlet component, it  
can hardly decay into $\tau$
leptons which is otherwise a standard search channel for MSSM heavy Higgs searches
\cite{jeremie2}. Here, we see that $H_2$
dominantly decays into the gauge bosons.
A considerable amount of efforts have already been put forward to discover BSM
Higgs scalars through its decay 
into the vector bosons modes at the LHC \cite{atlas1,atlas2,atlas3,atlas4,cms1} and
non observation of any new physics would eventually lead to the
upper bounds (UB) on $({\sigma\times Br})_{BSM Higgs}$ at 95\% CL. Using Ref:\cite{atlas3,atlas4} which assumes $\sqrt{s}=13~$TeV and 
integrated luminosity 
$\mathcal L = 36.1~fb^{-1}$,
we study the 
status of these BPs  
focusing on the promising $WW$ and $ZZ$ final states and the results have 
been displayed
in Table:\ref{table3}. Here the production 
of $H_2$ in association with the $b$ quark has not been considered where one may find upper limits on $(\sigma \times Br)_{BSM~Higgs}$ 
assuming $H_2$ decaying 
into $\tau\tau$ \cite{CMS_bbH_tautau} and $b\bar b$ \cite{CMS_bbH_bb} at $\sqrt{s}=13~$TeV run of the LHC.
In Table:\ref{table3}, different 
upper bounds $({\sigma\times Br})^{exp-UB}_{BSM Higgs}$ for the ggF and the VBF 
modes have also been shown respectively. Here one assumes $WW$
decaying leptonically to $e\nu \mu\nu$.
Similarly, for $ZZ$ final states, $4l$ and $2l2\nu$ channels
are combined. In the last column the necessary 
improvements for the future runs of the LHC to probe our BPs in terms of the
 predicted 
${({\sigma\times Br})_{NMSSM}}$ have been calculated.
Clearly, the $ZZ$ final state has been appeared as the best
channel for observing the $H_2$ scalar at the future runs of the LHC. In fact the 
relatively
light Higgses with $m_{H_2}$(=250,500~GeV) in BP-A and in BP-B can be probed via $ZZ$ final state with 
roughly an order of magnitude enhancement in the experimental sensitivity 
while somewhat less enhancement 
would be necessary to probe $m_{H_2}=700$~GeV in the case of BP-C.
Additionally, in the latter case, one may also observe that the
ggF contribution to $H_2$ production becomes comparable with the VBF production channel though the latter still dominates. 
This is because for this parameter point $C^2_{H_2gg}$ is somewhat larger
compared to the previous BPs which can again be attributed to somewhat larger
$S_{2,2}$. Thus BP-C serves as a good example where the
ggF production mode can compete with
the VBF production channel. Here one may observe that
a roughly equal amount of 
enhancement in sensitivity 
would be required for both production channels to discover the Higgs boson at the future runs of the LHC. In the next section, we would study the prospects of 
observing this Higgs scalar at the high luminosity run of the LHC at
$\sqrt s =14$~TeV and at a future $pp$ collider with 27~TeV center of mass energy 
considering
the fact that the rate of production of the said scalar is dominated by the VBF channel.

\begin{table}[!htbp]
\begin{center}\
\begin{tabular}{|c|c|c|c|c|c|c|c|}
\hline
\hline
BP-A &{$H^{\rm ggF}_2\rightarrow FF$~(pb)} 
&{$H^{\rm VBF}_2\rightarrow FF$~(pb)}&
${\sigma \times Br}^{exp-UB}_{\tiny{BSM Higgs}}$(pb)& 
\multicolumn{2}{c|}{$\frac{{\sigma \times Br}^{exp-UB}_{\tiny{BSM Higgs}}}{{\sigma\times Br}_{NMSSM}}$}\\
\hline
\hline
$FF$& ${\sigma\times Br}_{NMSSM}$& ${\sigma\times Br}_{NMSSM}$& ({95\%CL})&{ggF}&{VBF}\\
\hline
\hline
$WW$ 
 & $1.1\times10^{-3}$&  $1.03\times10^{-2}$ &6.0,1.0&$6\times10^{3}$ 
&$1\times10^{2}$\\
\hline
$ZZ$ 
 & $4.6\times10^{-4}$&  $4.4\times10^{-3}$ &0.2,0.15&
$4.0\times10^{2}$&30\\
\hline
\hline
BP-B &{} 
&{}& {}&{}&{}\\
\hline
\hline
$WW$ 
 &$2.0\times10^{-4}$  & $1.4\times10^{-3}$  &0.6,0.3&
$3\times10^{3}$&$2\times10^{2}$\\
\hline
$ZZ$ 
 & $0.92\times10^{-4}$&  $0.7\times10^{-3}$ &0.06,0.04
&$6\times10^{2}$&50\\
\hline
\hline
BP-C&{} 
&{}& {}&{}&{}\\
\hline
\hline
$WW$ 
&$1.7\times10^{-3}$  & $3.6\times10^{-3}$ &0.25,~0.1 
&$1\times10^{2}$&30\\
\hline
$ZZ$ 
 & $0.8\times10^{-3}$&  $1.7\times10^{-3}$ &0.03,~0.02&40&10\\
\hline
\hline
\end{tabular}
\end{center}
\vskip 0.8cm
\caption{ NMSSM prediction 
${\sigma\times Br}_{NMSSM}$ for the BPs in Table:\ref{tab1} along with the
95\%CL upper bounds assuming $H_2$ decaying into the $WW/ZZ$ final states.
Here we assume $\sqrt{s}=13~$TeV run of the LHC and 
$\mathcal L = 36.1~fb^{-1}$\cite{atlas3,atlas4}. 
In the last column the necessary 
improvements for the future runs of LHC to probe the BPs in terms of the 
predicted 
${({\sigma\times Br})_{NMSSM}}$ through the ggF and the VBF production modes 
have been presented.
}
\label{table3}
\end{table}

So far we discuss about a BSM Higgs scalar which is primarily singlet dominated with non-negligible $H_d$
components in it. It is the lightest of the two BSM Higgs scalars in the present
context. We will see that the problem of discovering $H_2$ 
persists as a motivation for experimental searches. In this context, it may be interesting to know the status of 
other Higgs states. (i) Heavier CP-even Higgs state:
The heavier one is mostly doublet dominated 
(approximately MSSM-like) whose properties can be seen from the 
Table:\ref{tab4}. The effective production cross-section is completely 
dominated by the ggH channel through $t$ quark loop and has been calculated at NNLO with
SusHi \cite{sushi}.
In the shown BPs, lightest $H_3$ can be obtained for BP-A, $\sigma^{ggH_3}_{NMSSM}$ is approximately 1$pb$.
Afterwards, it also dominantly
decays into $t\bar t$ final states and non-negligibly to $H_1 H_2$ or $A_1 Z$ 
(when kinematically allowed). In case of $t\bar t$ final states, using the 
simplified analysis as carried out in ref:\cite{jeremie2}, it may be 
possible to discover the heavy MSSM-like Higgs $H_3$ at $\sqrt s=$~14 TeV LHC run with 300 $fb^{-1}$ data. Additionally, 
heavier Higgs to lighter Higgs decays could 
be an important search channel in this case. In fact Higgs$\rightarrow$ Higgs
decays can be an important search channel for the lighter BSM Higgs scalar as well
(see Table:\ref{tab1}). 
These 
can in turn lead to different
possibilities in the final states with additional jets in case of the 
Higgs productions through the VBF process. For the decay modes $H_1H_2$ and $H_1H_1$, stringent 
constraints can be obtained from the resonant SM Higgs pair productions 
where Higgs can subsequently decay into
into $b \bar b b \bar b$, $b \bar b \tau \tau$, $b \bar b \gamma \gamma$ and
$b \bar b l \nu l \nu$ \cite{atlashiggspairprod,cmshiggspairprod}.  
The effective production cross-section for
$pp\rightarrow H_2(H_3) \rightarrow H_1H_1 (H_1H_2)$ is found to be well below 
the present bounds
as obtained by $\sqrt{s}=13~$TeV LHC run with
$\mathcal L = 35.9~fb^{-1}$ (see e.g., the summary plot in ref:\cite{cmscurrent}).
(ii) CP-odd Higgs state: 
The lightest CP-odd Higgs is dominantly singlet-like while the heavier
one is doublet-like. Their masses have been shown in Table:\ref{tab4}.
The production of the CP-odd states would be dominated
by the gluon-gluon fusion process. Clearly, the coupling of relevance would be $A_igg$. For the lightest CP-odd Higgs scalar, we find that $A_1gg$
becomes maximum for BP-A and the reduced coupling becomes
$C^2_{A_1 gg} \sim 0.1$. Similarly, for the same BP, one may read 
$C^2_{A_2 gg} \sim 0.36$. In case of other BPs, one gets somewhat smaller values for 
$C^2_{A_2 gg}$. Thus BP-A may potentially be seen as best example to search for the CP-odd states at the LHC in the present context. 
After production,
for BP-A and BP-B, 
the CP-odd scalars $A_1$ and $A_2$ dominantly decay in to $t\bar t$ 
with branching ratio more than $90\%$. In fact the MSSM-like CP-odd Higgs state
$A_2$ may be probed at $\sqrt s=$~14 TeV LHC run with 300 $fb^{-1}$ data just
like the MSSM-like heavy CP even Higgs state $H_3$ \cite{jeremie2}.
In case of BP-C, $A_2$ can decay in to 
$Z H_2$ with 20\% BR and $A_1 H_1$ with 10\% BR. However, with
$m_{A_2}\sim 1$~TeV, the production rate would be suppressed to search it through the cascades of lighter Higgs scalars.
\vskip 0.5cm
\begin{table}[!htbp]
\begin{center}
\begin{tabular}[vt]{clll}
\hline
\hline
parameter & ~~A &~~ B & ~~C  \\
\hline
\hline
$m_{H_{3}}$ ~(GeV) &~~524&~~762.1&~~1017.1\\
\hline
\hline
$m_{A_1}$~(GeV) &~~480.8&~~627.9&~~809.0 \\
$m_{A_{2}}$ ~(GeV)&~~524.9&~~763.6&~~1035.0 \\
\hline
\hline
$S_{3,1}$ & ~~0.87 &~~0.87&~~ 0.85 \\
$S_{3,2}$ & ~~-0.46 &~~-0.46 &~~ -0.44 \\
$S_{3,3}$ & ~~-0.17 &~~-0.16&~~ -0.27 \\
\hline
\hline
$C^2_{H_3 t\bar t}$ &~~0.265 &~~0.26 & ~~0.25  \\
$C^2_{H_3 b\bar b}$ &~~ 3.77&~~ 3.80 & ~~3.58  \\
$C^2_{H_3 \tau\bar \tau}$ &~~ 3.78&~~ 3.82 & ~~3.62  \\
$C^2_{H_3 WW/H_3ZZ}$ &~~ 0.00053&~~ 0.0003 & ~~0.0004  \\
$C^2_{H_3 gg}$ &~~ 0.268&~~ 0.26 & ~~0.25   \\
\hline
\hline
$\sigma^{ggH_3}_{NMSSM}$~(pb)~&~~ 0.982 &~~0.16 &~~0.028 \\
\hline
\hline
$Br({H_3\rightarrow t\bar t})$ &~~ 0.86&~~0.9 & ~~0.65 \\
$Br({H_3\rightarrow H_1 H_2})$ 
&~~~0.12 &~~0.07&~~0.15 \\
$Br({H_3\rightarrow A_1 Z})$ 
&~~~- &~0.012&~~0.1 \\
\hline
\hline
\end{tabular}
\vskip 0.8cm
\caption{Mass, compositions,
reduced couplings, cross-sections and leading Brs for the 
heavier MSSM-like Higgs scalar $H_3$.}
\label{tab4}
\end{center}
\end{table}

\section{LHC phenomenology at $\sqrt s =14$ and $\sqrt s =27$~TeV}
\label{sec:simulation}
In this section we would study the search prospects of the lighter BSM
Higgs bosons for the aforesaid benchmark points at $\sqrt s =14$~TeV and $\sqrt s =27$~TeV run of the LHC. In particular we would consider production of the
$H_2$ scalar 
through VBF and its decay $H_2 \rightarrow ZZ$ (as it is found to be the most promising channel) where $Z$ can decay into leptons, $Z \rightarrow l^+l^-$ (where $l=e,\mu$).
One can characterize the VBF induced processes, $pp \rightarrow jjH_2$ ($j$ refers to light jets) by
the presence of two energetic jets with a large rapidity gap. For a detailed collider simulation, we generate parton level events using MadGraph5 with
PDF choice NN23lo1
\cite{Alwall:2011uj,Alwall:2014hca,Ball:2012cx,Ball:2014uwa}
and subsequently passed the events into Pythia (v8) \cite {Sjostrand:2006za}
for hadronization taking into account the initial state
radiation/final state radiation (ISR/FSR) and
multiple interactions. We use MadGraph5 for implementing decays of the
lightest BSM Higgs scalar. The jets, leptons, photons
have been reconstructed using fast detector simulator Delphes-v3.3.3 \cite{deFavereau:2013fsa,Selvaggi:2014mya,Mertens:2015kba}. The jets and leptons are reconstructed by anti-$k_t$
algorithm \cite{Cacciari:2008gp} implemented in the Fastjet
\cite{Cacciari:2005hq,Cacciari:2008gp,Cacciari:2011ma} with a cone of $\Delta R =0.4$ and minimum transverse momentum of 20 GeV.
We generate our new physics
signal events associated with $H_2$ production at the
leading order (LO) using the UFO file based on Ref:\cite{Maltoni:2013sma} which
considers VBF at NLO-QCD accuracy.
Since signal does not include much missing 
transverse energy, the dominant background appears to be 
$pp \rightarrow ZZ + 2j$ (where
j stands for light jets) where $Z$ decays leptonically \cite{atlas4}. There may be
other sub-dominant background processes like $pp \rightarrow \gamma \gamma + 2j$, $pp \rightarrow \gamma Z + 2j$ or $pp \rightarrow W Z + 2j$
.  Then we use the following selection cuts \cite{atlas4} to optimize the 
significance ($\mathcal S = {S \over \sqrt{B + S}}$). 
\begin{itemize}
\item
We demand the presence of two same-flavour, opposite-sign leptons (electron or muon)
with $p^{e}_{T} $ or $p^{\mu}_{T} > 15$~GeV. This criterion is somewhat
different compared to what has been assumed in Ref:\cite{atlas4} where
the highest-$p_T$
lepton in the quadruplet must satisfy
$p_T >20$ GeV, and the second (third) lepton must satisfy
$p_T>15$ GeV (10 GeV). Similarly, the pseudorapidity range of
electron or muon has been considered to lie $|\eta| < 2.47 ~(2.7)$.
\item
  We require the two leading forward jets to satisfy $p_{T} \ge 30$~GeV and $|\eta| < 4.5$
  . Moreover, these two jets should lie in opposite hemispheres with $\eta_{j_1} \times \eta_{j_2} \le$ 0.
\item
Forward jets should also satisfy 
$|\Delta \eta_{j_{1}j_{2}}| \ge$ 3.3 and di-jet invariant mass, 
$m_{j_{1}j_{2}} \ge 400$~GeV \cite{atlas4}.

\item While computing the statistical significance, we impose additional cuts : (i)
  For BP-A ($m_{H_2}=250$~GeV), the four lepton invariant mass ($m_{4l}$) should lie between
  $240-260~$GeV and (ii) for BP-C ($m_{H_2}=700$~GeV), $m_{4l}$ should be above $600~$GeV.
\end{itemize}
For background simulation, we generate $pp \rightarrow ZZ$ in 
association with 0-2jets at the matrix element level using MadGraph5. Here we also implement the MLM matching technique to avoid double counting. The production may arise from quark–antiquark annihilation, gluon-initiated production associated with jets and from EW vector-boson
scattering. The last process plays a more important role in the VBF-enriched category which makes signal discrimination very difficult from the backgrounds. Considering the fact that effective production cross-section of the signal process
$pp\rightarrow H_2 jj \rightarrow ZZ jj \rightarrow l^+l^-l^+l^- jj$ is not quite large for singlet like Higgs, we find that probing lighter BSM Higgs with
mass $m_{H_2}=250$ GeV (BP-A) is more difficult even at the upgraded LHC. In order to have better significance we devise a small window of 20~GeV around the Higgs boson mass in the first case.

For quantitative estimates,
we compute the number of signal and background events
for the lighter and heavier BSM Higgs scalar for the BP-A and BP-C. 
As shown in Table:\ref{table3}, for these BPs, 
the necessary 
improvements which would be required to probe them at the future runs of LHC 
would be smaller compared to the BP-B. This is related to smaller $C^2_{H_2ZZ}$ which is in turn related to smaller $S_{2,1}$ in case of BP-B.
The production cross-section
for the $H_2$ scalar has been evaluated \cite{Maltoni:2013sma} at the NLO-QCD level which resembles quite well with \cite{Hussein:2017pjz} while for the background  productions
$pp \rightarrow ZZ jj \rightarrow l^+l^-l^+l^- jj $ ($l=e,\mu$), we use
LO estimates multiplied by K-factor. At the leading order
our computation produces $\sigma({pp\rightarrow ZZ jj})= 17.4\times10^{3}$ fb 
at the $\sqrt s =14$~TeV and $44.3\times10^{3}$ fb for the $\sqrt s = 27$~TeV 
runs of the LHC. In
Ref:\cite{Sirunyan:2017zjc,Aaboud:2017rwm}, the total $ZZ$ cross-section predictions is computed to be 18~pb with MATRIX \cite{Cascioli:2014yka} at NNLO in QCD for $\sqrt s =14$~TeV which is quite consistent with our result. 
We use a conservative choice for K-factor ($\sim$ 1.0) which is consistent
with the limit \cite{Campanario:2014ioa} if one considers the
scale uncertainty.
We note that for the 27 TeV run of the LHC, we assume the detector acceptance and efficiency are the same as the 14 TeV run.  
In Table \ref{table14tev-cross} and \ref{tablecolli}, we show the effective production cross-section for signal ($\sigma_{}^{eff} = \sigma({pp\rightarrow H_2jj\rightarrow ZZ jj\rightarrow 4ljj})\times efficiency$) and for the backgrounds
($\sigma_{B}^{eff} = \sigma({pp\rightarrow ZZ jj\rightarrow 4ljj})\times efficiency$) and the number of
typical signal and background events obtained
for 3 and 15 ${\rm ab}^{-1}$ 
integrated luminosity ($\mathcal L$) for $\sqrt s =14$~ and 27~TeV runs of the LHC respectively. One may see that the effective production cross-section 
for signal events $\sigma_{}^{eff}$ is not quite large, though the situation 
somewhat improves for higher cms energy. This is in general responsible for
having a small number of signal events at the LHC.
Hence, though the Higgs production through the VBF mode appears to be an interesting possibility, but it really lacks to provide with the desired sensitivity at the high-luminosity run or even
at the upgraded LHC. This is especially true when the lightest 
non SM-like Higgs scalar is light. 
Situation improves for a heavier Higgs boson, thanks to the rapid fall of the four lepton backgrounds with its invariant mass.

\begin{table}[!htb]\centering
\begin{tabular}{|c|c|cc|}
\hline
$\sqrt{s}$ (TeV)& &A & C  \\
\hline

\hline
14&$\sigma^{pp\rightarrow H_2jj}$,$\sigma_{}^{eff}$ (fb) &1818, 0.00062 
& 242, $0.0004$ \\
\hline
27& $\sigma^{pp\rightarrow H_2jj}$,$\sigma_{}^{eff}$(fb) &5856, 0.002 
& 1135, $0.002$ \\
\hline 
14&$\sigma_{B}^{eff}$(fb) & $0.006$
 & $0.004$\\
\hline 
27&$\sigma_{B}^{eff}$(fb) & $0.016$
& $0.010$\\
\hline 
\end{tabular}
\caption{ VBF and Effective production cross-section for the lighter and heavier  Higgs scalars
  (BPs A\&C). For backgrounds, we calculated $\sigma({pp\rightarrow ZZ jj})= 18\times10^{3}$ and $45.6\times10^{3}$
  fb for $\sqrt s =14,27$~TeV runs of the LHC respectively which leads to the said $\sigma_{B}^{eff}$.
  }
\label{table14tev-cross}
\end{table}

\begin{table}[!htb]\centering
\begin{tabular}{|c|cc|}
\hline
& A ($\sqrt s =14,27$ TeV) & C ($\sqrt s =14,27$ TeV) \\
\hline
Signal (S) & 2, 30   & 1.2, 30\\
\hline 
Backgrounds (B) & 18,240   &12,150  \\
\hline 
Significance ($\mathcal S = {S \over \sqrt{B + S}}$) & 0.4,1.8 &0.3, 2.2  \\
\hline 
\end{tabular}
\caption{Expected number of Signal and background events that qualify the selection criteria, mentioned in the text for the BPs A and C. 
We take ${\mathcal L}=3 \& 15~{\rm ab}^{-1}$ integrated luminosity for
the $\sqrt s =14$ and 27~TeV runs of the LHC respectively.
}
\label{tablecolli}
\end{table}

\section{Conclusion}
\label{sec:con}
In this work, we show that the rate of non SM-like
Higgs productions via the
gluon-gluon fusion process could be insignificant 
in some of parts of the NMSSM parameter space. This can be attributed to the smallness
of up-type Higgs component in the Higgs scalar and also to the small values 
of $\tan\beta$.  
In this case, the vector-boson fusion 
may play the
leading role in the BSM Higgs searches if the new state is not completely dominated by singlet component. We present three benchmark points where 
the mass of the non SM-like Higgs scalar may vary from 250~GeV to 700~GeV. Then
we study
the role of existing LHC constraints on the said BPs 
in the light of the 13 TeV data. We also  
discuss the prospects of observing such a state at the high-luminosity 
${\mathcal L}=3~{\rm ab}^{-1}$ run of the LHC at cms energy 
$\sqrt s =14$~TeV and at 
the proposed $\sqrt s =27$~TeV run of the LHC with ${\mathcal L}=15 ~{\rm ab}^{-1}$. The dominant background comes from $pp \rightarrow ZZ +2jets$ in which VBF  may also play a significant role which 
makes it somewhat difficult to distinguish the backgrounds from the signal, especially for lighter BSM Higgs scalar in BP-A ($m_{H_2}=250$~GeV).
For heavier BSM Higgs, we can use a stronger cut on the 4 lepton
invariant mass, $m_{4l}$ and in turn this can yield better significance. Finally, in view of this work, it may be interesting to study the prospects of
such a BSM Higgs scalar at the Linear Colliders.

\section*{Acknowledgments}
I am indebted to Ulrich Ellwanger and Amit Chakraborty for reading the draft and their suggestions. I am thankful to AseshKrishna Datta and Kirtiman Ghosh for discussions.



\begin{thebibliography}{999}
\bibitem{Aad:2015zhl} 
  G.~Aad {\it et al.} [ATLAS and CMS Collaborations],
  Phys.\ Rev.\ Lett.\  {\bf 114}, 191803 (2015)
  doi:10.1103/PhysRevLett.114.191803
  [arXiv:1503.07589 [hep-ex]].

\bibitem{HiggsDiscoveryJuly2012}
 G.~Aad {\it et al.}  [ATLAS Collaboration],
  Phys.\ Lett.\ B {\bf 716}, 1 (2012)
  [arXiv:1207.7214 [hep-ex]];\\
S.~Chatrchyan {\it et al.}  [CMS Collaboration],
  Phys.\ Lett.\ B {\bf 716}, 30 (2012)
  [arXiv:1207.7235 [hep-ex]].

\bibitem{SUSYreviews1}
For reviews on supersymmetry, see, { e.g.},
H. P. xNilles, \PREP(110, 1, 1984);
J.~D.~Lykken,
  hep-th/9612114;
J. Wess and J. Bagger, {\it Supersymmetry and Supergravity}, 2nd ed.,
(Princeton, 1991);
  D.~J.~H.~Chung, L.~L.~Everett, G.~L.~Kane, S.~F.~King,
J.~D.~Lykken and L.~T.~Wang,
  Phys.\ Rept.\  {\bf 407}, 1 (2005);
H. E. Haber and G. Kane, \PREP(117, 75, 1985) ;
S.~P.~Martin,
arXiv:hep-ph/9709356.
\bibitem{SUSYbook1}
M. Drees, P. Roy and R. M. Godbole,
{\it Theory and Phenomenology of Sparticles},
(World Scientific, Singapore, 2005);  
  H.~Baer and X.~Tata,
  {\it{Weak scale supersymmetry: From superfields to scattering events}},
  Cambridge, UK: Univ. Pr. (2006) 537 p.

\bibitem{djouadi}
 A.~Djouadi,
  Phys.\ Rept.\  {\bf 459}, 1 (2008)
  [hep-ph/0503173].

\bibitem{ccb} 
  J.~E.~Camargo-Molina, B.~O'Leary, W.~Porod and F.~Staub,
  JHEP {\bf 1312}, 103 (2013)
  doi:10.1007/JHEP12(2013)103
  [arXiv:1309.7212 [hep-ph]];
  N.~Blinov and D.~E.~Morrissey,
  JHEP {\bf 1403}, 106 (2014)
  doi:10.1007/JHEP03(2014)106
  [arXiv:1310.4174 [hep-ph]];
  D.~Chowdhury, R.~M.~Godbole, K.~A.~Mohan and S.~K.~Vempati,
  JHEP {\bf 1402}, 110 (2014)
  doi:10.1007/JHEP02(2014)110
  [arXiv:1310.1932 [hep-ph]];
  J.~E.~Camargo-Molina, B.~Garbrecht, B.~O'Leary, W.~Porod and F.~Staub,
  Phys.\ Lett.\ B {\bf 737}, 156 (2014)
  doi:10.1016/j.physletb.2014.08.036
  [arXiv:1405.7376 [hep-ph]];
 U.~Chattopadhyay and A.~Dey,
  JHEP {\bf 1411}, 161 (2014)
  doi:10.1007/JHEP11(2014)161
  [arXiv:1409.0611 [hep-ph]].


\bibitem{FT}
  R.~Barbieri and G.~F.~Giudice,
  Nucl.\ Phys.\ B {\bf 306}, 63 (1988);
J.~R.~Ellis, K.~Enqvist, D.~V.~Nanopoulos and F.~Zwirner,
  Mod.\ Phys.\ Lett.\ A {\bf 1}, 57 (1986);
  R.~Kitano and Y.~Nomura,
  Phys.\ Lett.\ B {\bf 631}, 58 (2005)
  [hep-ph/0509039];
  M.~E.~Cabrera, J.~A.~Casas and R.~Ruiz de Austri,
  JHEP {\bf 0903}, 075 (2009)
  [arXiv:0812.0536 [hep-ph]];
  D.~M.~Ghilencea,
  PoS Corfu {\bf 2012}, 034 (2013)
  [arXiv:1304.1193 [hep-ph]];
  D.~M.~Ghilencea,
  Nucl.\ Phys.\ B {\bf 876}, 16 (2013)
  [arXiv:1302.5262 [hep-ph]];
  D.~M.~Ghilencea and G.~G.~Ross,
  Nucl.\ Phys.\ B {\bf 868}, 65 (2013)
  [arXiv:1208.0837 [hep-ph]];
   H.~Baer, V.~Barger, P.~Huang, A.~Mustafayev and X.~Tata,
  Phys.\ Rev.\ Lett.\  {\bf 109}, 161802 (2012)
  [arXiv:1207.3343 [hep-ph]];
H.~Baer, V.~Barger and D.~Mickelson,
Phys.\ Rev.\ D {\bf 88}, no. 9, 095013 (2013)
[arXiv:1309.2984 [hep-ph]];
M.~W.~Cahill-Rowley, J.~L.~Hewett, A.~Ismail and T.~G.~Rizzo,
Phys.\ Rev.\ D {\bf 86}, 075015 (2012)
[arXiv:1206.5800 [hep-ph]];
M.~W.~Cahill-Rowley, J.~L.~Hewett, A.~Ismail and T.~G.~Rizzo,
Phys.\ Rev.\ D {\bf 88}, no. 3, 035002 (2013)
[arXiv:1211.1981 [hep-ph]].
M.~Cahill-Rowley, R.~Cotta, A.~Drlica-Wagner, S.~Funk, J.~Hewett, A.~Ismail, T.~Rizzo and M.~Wood,
Phys.\ Rev.\ D {\bf 91}, no. 5, 055011 (2015)
[arXiv:1405.6716 [hep-ph]];
M.~Perelstein and C.~Spethmann,
  JHEP {\bf 0704}, 070 (2007)
  [hep-ph/0702038];
  C.~Boehm, P.~S.~B.~Dev, A.~Mazumdar and E.~Pukartas,
  JHEP {\bf 1306}, 113 (2013)
  [arXiv:1303.5386 [hep-ph]];
K.~L.~Chan, U.~Chattopadhyay and P.~Nath,
Phys.\ Rev.\ D {\bf 58}, 096004 (1998).

\bibitem{Branco} 
  G.~C.~Branco, P.~M.~Ferreira, L.~Lavoura, M.~N.~Rebelo, M.~Sher and J.~P.~Silva,
  Phys.\ Rept.\  {\bf 516}, 1 (2012)
  doi:10.1016/j.physrep.2012.02.002
  [arXiv:1106.0034 [hep-ph]].

\bibitem{Zarikas:1995qb} 
  V.~Zarikas,
  Phys.\ Lett.\ B {\bf 384}, 180 (1996)
  doi:10.1016/0370-2693(96)00701-0
  [hep-ph/9509338].
\bibitem{Lahanas:1998wf} 
  A.~B.~Lahanas, V.~C.~Spanos and V.~Zarikas,
  Phys.\ Lett.\ B {\bf 472}, 119 (2000)
  doi:10.1016/S0370-2693(99)01400-8
  [hep-ph/9812535].
\bibitem{Aliferis:2014ofa} 
  G.~Aliferis, G.~Kofinas and V.~Zarikas,
  Phys.\ Rev.\ D {\bf 91}, no. 4, 045002 (2015)
  doi:10.1103/PhysRevD.91.045002
  [arXiv:1406.6215 [hep-ph]].


\bibitem{review1} 
  M.~Maniatis,
  Int.\ J.\ Mod.\ Phys.\ A {\bf 25}, 3505 (2010)
  doi:10.1142/S0217751X10049827
  [arXiv:0906.0777 [hep-ph]].
\bibitem{review2}
  U.~Ellwanger, C.~Hugonie and A.~M.~Teixeira,
  Phys.\ Rept.\  {\bf 496}, 1 (2010)
  doi:10.1016/j.physrep.2010.07.001
  [arXiv:0910.1785 [hep-ph]].

\bibitem{higgs1} 
  U.~Ellwanger and C.~Hugonie,
  Mod.\ Phys.\ Lett.\ A {\bf 22}, 1581 (2007)
  doi:10.1142/S0217732307023870
  [hep-ph/0612133].

\bibitem{higgs2} 
  U.~Ellwanger,
  Eur.\ Phys.\ J.\ C {\bf 71}, 1782 (2011)
  doi:10.1140/epjc/s10052-011-1782-3
  [arXiv:1108.0157 [hep-ph]].


 \bibitem{nmssmft} 
  M.~Bastero-Gil, C.~Hugonie, S.~F.~King, D.~P.~Roy and S.~Vempati,
  Phys.\ Lett.\ B {\bf 489}, 359 (2000)
  doi:10.1016/S0370-2693(00)00930-8
  [hep-ph/0006198];
  R.~Dermisek and J.~F.~Gunion,
  Phys.\ Rev.\ D {\bf 73}, 111701 (2006)
  doi:10.1103/PhysRevD.73.111701
  [hep-ph/0510322];
  U.~Ellwanger, G.~Espitalier-Noel and C.~Hugonie,
  JHEP {\bf 1109}, 105 (2011)
  doi:10.1007/JHEP09(2011)105
  [arXiv:1107.2472 [hep-ph]];
  G.~G.~Ross and K.~Schmidt-Hoberg,
  Nucl.\ Phys.\ B {\bf 862}, 710 (2012)
  doi:10.1016/j.nuclphysb.2012.05.007
  [arXiv:1108.1284 [hep-ph]];
  G.~G.~Ross, K.~Schmidt-Hoberg and F.~Staub,
  JHEP {\bf 1208}, 074 (2012)
  doi:10.1007/JHEP08(2012)074
  [arXiv:1205.1509 [hep-ph]];
  T.~Gherghetta, B.~von Harling, A.~D.~Medina and M.~A.~Schmidt,
  JHEP {\bf 1302}, 032 (2013)
  doi:10.1007/JHEP02(2013)032
  [arXiv:1212.5243 [hep-ph]];
  M.~Perelstein and B.~Shakya,
  Phys.\ Rev.\ D {\bf 88}, no. 7, 075003 (2013)
  doi:10.1103/PhysRevD.88.075003
  [arXiv:1208.0833 [hep-ph]];
  D.~Kim, P.~Athron, C.~Balázs, B.~Farmer and E.~Hutchison,
  Phys.\ Rev.\ D {\bf 90}, no. 5, 055008 (2014)
  doi:10.1103/PhysRevD.90.055008
  [arXiv:1312.4150 [hep-ph]];
  A.~Kaminska, G.~G.~Ross, K.~Schmidt-Hoberg and F.~Staub,
  JHEP {\bf 1406}, 153 (2014)
  doi:10.1007/JHEP06(2014)153
  [arXiv:1401.1816 [hep-ph]];
  M.~Y.~Binjonaid and S.~F.~King,
  Phys.\ Rev.\ D {\bf 90}, no. 5, 055020 (2014)
  Erratum: [Phys.\ Rev.\ D {\bf 90}, no. 7, 079903 (2014)]
  doi:10.1103/PhysRevD.90.079903, 10.1103/PhysRevD.90.055020
  [arXiv:1403.2088 [hep-ph]].
\bibitem{Cao}
J.~J.~Cao, Z.~X.~Heng, J.~M.~Yang, Y.~M.~Zhang and J.~Y.~Zhu,
and NMSSM,''
  JHEP {\bf 1203}, 086 (2012)
  doi:10.1007/JHEP03(2012)086
  [arXiv:1202.5821 [hep-ph]];
 J.~Cao, Y.~He, L.~Shang, W.~Su and Y.~Zhang,
  JHEP {\bf 1608}, 037 (2016)
  doi:10.1007/JHEP08(2016)037
  [arXiv:1606.04416 [hep-ph]];
 J.~Cao, F.~Ding, C.~Han, J.~M.~Yang and J.~Zhu,
  JHEP {\bf 1311}, 018 (2013)
  doi:10.1007/JHEP11(2013)018
  [arXiv:1309.4939 [hep-ph]];
J.~Cao and J.~M.~Yang,
  Phys.\ Rev.\ D {\bf 78}, 115001 (2008)
  doi:10.1103/PhysRevD.78.115001
  [arXiv:0810.0989 [hep-ph]];
J.~Cao, D.~Li, L.~Shang, P.~Wu and Y.~Zhang,
Pair Production at the LHC,''
  JHEP {\bf 1412}, 026 (2014)
  doi:10.1007/JHEP12(2014)026
  [arXiv:1409.8431 [hep-ph]].
\bibitem{nmssmhiggslhc1} 
  H.~K.~Dreiner, F.~Staub and A.~Vicente,
  Phys.\ Rev.\ D {\bf 87}, no. 3, 035009 (2013)
  doi:10.1103/PhysRevD.87.035009
  [arXiv:1211.6987 [hep-ph]].
\bibitem{nmssmhiggslhc2} 
  S.~F.~King, M.~Mühlleitner, R.~Nevzorov and K.~Walz,
  Nucl.\ Phys.\ B {\bf 870}, 323 (2013)
  doi:10.1016/j.nuclphysb.2013.01.020
  [arXiv:1211.5074 [hep-ph]].


\bibitem{nmssmhiggslhc3} 
  Z.~Kang, J.~Li, T.~Li, D.~Liu and J.~Shu,
  Phys.\ Rev.\ D {\bf 88}, no. 1, 015006 (2013)
  doi:10.1103/PhysRevD.88.015006
  [arXiv:1301.0453 [hep-ph]].


\bibitem{nmssmhiggslhc4} 
  D.~G.~Cerdeno, P.~Ghosh and C.~B.~Park,
  JHEP {\bf 1306}, 031 (2013)
  doi:10.1007/JHEP06(2013)031
  [arXiv:1301.1325 [hep-ph]].


\bibitem{nmssmhiggslhc5} 
  D.~G.~Cerdeño, P.~Ghosh, C.~B.~Park and M.~Peiró,
  JHEP {\bf 1402}, 048 (2014)
  doi:10.1007/JHEP02(2014)048
  [arXiv:1307.7601 [hep-ph]].
\bibitem{nmssmhiggslhc6} 
  S.~F.~King, M.~Mühlleitner, R.~Nevzorov and K.~Walz,
  Phys.\ Rev.\ D {\bf 90}, no. 9, 095014 (2014)
  doi:10.1103/PhysRevD.90.095014
  [arXiv:1408.1120 [hep-ph]].
\bibitem{nmssmhiggslhc7} 
  A.~Chakraborty, D.~K.~Ghosh, S.~Mondal, S.~Poddar and D.~Sengupta,
  Phys.\ Rev.\ D {\bf 91}, 115018 (2015)
  doi:10.1103/PhysRevD.91.115018
  [arXiv:1503.07592 [hep-ph]].
\bibitem{Beuria1} 
  J.~Beuria, A.~Chatterjee, A.~Datta and S.~K.~Rai,
  JHEP {\bf 1509}, 073 (2015)
  doi:10.1007/JHEP09(2015)073
  [arXiv:1505.00604 [hep-ph]].


\bibitem{nmssmhiggslhc8} 
  F.~Staub, P.~Athron, U.~Ellwanger, R.~Gröber, M.~Mühlleitner, P.~Slavich and A.~Voigt,
  Comput.\ Phys.\ Commun.\  {\bf 202}, 113 (2016)
  doi:10.1016/j.cpc.2016.01.005
  [arXiv:1507.05093 [hep-ph]].
\bibitem{Guchait} 
  M.~Guchait and J.~Kumar,
  Int.\ J.\ Mod.\ Phys.\ A {\bf 31}, no. 12, 1650069 (2016)
  doi:10.1142/S0217751X1650069X
  [arXiv:1509.02452 [hep-ph]].

\bibitem{nmssmhiggslhc9} 
  U.~Ellwanger and M.~Rodriguez-Vazquez,
  JHEP {\bf 1602}, 096 (2016)
  doi:10.1007/JHEP02(2016)096
  [arXiv:1512.04281 [hep-ph]].
\bibitem{Beuria2} 
  J.~Beuria, A.~Chatterjee and A.~Datta,
  JHEP {\bf 1608}, 004 (2016)
  doi:10.1007/JHEP08(2016)004
  [arXiv:1603.08463 [hep-ph]].
\bibitem{Guchait2} 
  M.~Guchait and J.~Kumar,
  Phys.\ Rev.\ D {\bf 95}, no. 3, 035036 (2017)
  doi:10.1103/PhysRevD.95.035036
  [arXiv:1608.05693 [hep-ph]].


\bibitem{nmssmhiggslhc10} 
  S.~P.~Das and M.~Nowakowski,
  Phys.\ Rev.\ D {\bf 96}, no. 5, 055014 (2017)
  doi:10.1103/PhysRevD.96.055014
  [arXiv:1612.07241 [hep-ph]].

\bibitem{Baum1} 
  S.~Baum, K.~Freese, N.~R.~Shah and B.~Shakya,
  Phys.\ Rev.\ D {\bf 95}, no. 11, 115036 (2017)
  doi:10.1103/PhysRevD.95.115036
  [arXiv:1703.07800 [hep-ph]].
\bibitem{Das1} 
  B.~Das, S.~Moretti, S.~Munir and P.~Poulose,
  Eur.\ Phys.\ J.\ C {\bf 77}, no. 8, 544 (2017)
  doi:10.1140/epjc/s10052-017-5096-y
  [arXiv:1704.02941 [hep-ph]].


\bibitem{nmssmhiggslhc11} 
  U.~Ellwanger and M.~Rodriguez-Vazquez,
  JHEP {\bf 1711}, 008 (2017)
  doi:10.1007/JHEP11(2017)008
  [arXiv:1707.08522 [hep-ph]].


\bibitem{nmssmhiggslhc12} 
  S.~P.~Das, J.~Fraga and C.~Avila,
  arXiv:1712.04395 [hep-ph].

\bibitem{singlinonmssm1} 
  D.~Das, U.~Ellwanger and A.~M.~Teixeira,
  JHEP {\bf 1304}, 117 (2013)
  doi:10.1007/JHEP04(2013)117
  [arXiv:1301.7584 [hep-ph]].

\bibitem{singlinonmssm2} 
  D.~Das, U.~Ellwanger and A.~M.~Teixeira,
  JHEP {\bf 1204}, 067 (2012)
  doi:10.1007/JHEP04(2012)067
  [arXiv:1202.5244 [hep-ph]].
\bibitem{singlinonmssm3} 
  U.~Ellwanger and A.~M.~Teixeira,
  JHEP {\bf 1410}, 113 (2014)
  doi:10.1007/JHEP10(2014)113
  [arXiv:1406.7221 [hep-ph]].
\bibitem{LEP} 
  S.~Schael {\it et al.} [ALEPH and DELPHI and L3 and OPAL Collaborations and LEP Working Group for Higgs Boson Searches],
  Eur.\ Phys.\ J.\ C {\bf 47}, 547 (2006)
  doi:10.1140/epjc/s2006-02569-7
  [hep-ex/0602042].
\bibitem{magg1} 
  D.~T.~Nhung, M.~Muhlleitner, J.~Streicher and K.~Walz,
  JHEP {\bf 1311}, 181 (2013)
  doi:10.1007/JHEP11(2013)181
  [arXiv:1306.3926 [hep-ph]].
\bibitem{nmssmhiggs1} 
  J.~F.~Gunion, H.~E.~Haber and T.~Moroi,
  eConf C {\bf 960625}, LTH095 (1996)
  [hep-ph/9610337];
\bibitem{nmssmhiggs2} 
  R.~Dermisek and J.~F.~Gunion,
  Phys.\ Rev.\ D {\bf 73}, 111701 (2006)
  doi:10.1103/PhysRevD.73.111701
  [hep-ph/0510322];
  R.~Dermisek and J.~F.~Gunion,
  Phys.\ Rev.\ D {\bf 75}, 075019 (2007)
  doi:10.1103/PhysRevD.75.075019
  [hep-ph/0611142].



\bibitem{Djouadihiggs} 
  A.~Djouadi,
  Phys.\ Rept.\  {\bf 457}, 1 (2008)
  doi:10.1016/j.physrep.2007.10.004
  [hep-ph/0503172].


\bibitem{Reina:2005ae} 
  L.~Reina,
  hep-ph/0512377.



\bibitem{Ellishiggs} 
  J.~Ellis,
  arXiv:1702.05436 [hep-ph].

\bibitem{deFlorian:2016spz} 
  D.~de Florian {\it et al.} [LHC Higgs Cross Section Working Group],
  doi:10.23731/CYRM-2017-002
  arXiv:1610.07922 [hep-ph].

\bibitem{vbf}
  D.~R.~T.~Jones and S.~T.~Petcov,
  Phys.\ Lett.\  {\bf 84B} (1979) 440.
  doi:10.1016/0370-2693(79)91234-6

\bibitem{jeremie1} 
  A.~Djouadi and J.~Quevillon,
  JHEP {\bf 1310}, 028 (2013)
  doi:10.1007/JHEP10(2013)028
  [arXiv:1304.1787 [hep-ph]].

\bibitem{jeremie2} 
  A.~Djouadi, L.~Maiani, A.~Polosa, J.~Quevillon and V.~Riquer,
  JHEP {\bf 1506}, 168 (2015)
  doi:10.1007/JHEP06(2015)168
  [arXiv:1502.05653 [hep-ph]].

\bibitem{fermio}
H.~Pois, T.~J.~Weiler and T.~C.~Yuan,
  Phys.\ Rev.\ D {\bf 47} (1993) 3886
  doi:10.1103/PhysRevD.47.3886
  [hep-ph/9303277];
A.~Stange, W.~J.~Marciano and S.~Willenbrock,
  Phys.\ Rev.\ D {\bf 49} (1994) 1354
  doi:10.1103/PhysRevD.49.1354
  [hep-ph/9309294];
A.~G.~Akeroyd,
  Phys.\ Lett.\ B {\bf 368} (1996) 89
  doi:10.1016/0370-2693(95)01478-0
  [hep-ph/9511347];
A.~G.~Akeroyd,
  J.\ Phys.\ G {\bf 24} (1998) 1983
  doi:10.1088/0954-3899/24/11/001
  [hep-ph/9803324];
A.~Barroso, L.~Brucher and R.~Santos,
  Phys.\ Rev.\ D {\bf 60} (1999) 035005
  doi:10.1103/PhysRevD.60.035005
  [hep-ph/9901293];
L.~Brucher and R.~Santos,
  Eur.\ Phys.\ J.\ C {\bf 12} (2000) 87
  doi:10.1007/s100529900252
  [hep-ph/9907434];
G.~L.~Landsberg and K.~T.~Matchev,
  Phys.\ Rev.\ D {\bf 62} (2000) 035004
  doi:10.1103/PhysRevD.62.035004
  [hep-ex/0001007];
A.~G.~Akeroyd and M.~A.~Diaz,
  Phys.\ Rev.\ D {\bf 67} (2003) 095007
  doi:10.1103/PhysRevD.67.095007
  [hep-ph/0301203];
A.~G.~Akeroyd, A.~Alves, M.~A.~Diaz and O.~J.~P.~Eboli,
  Eur.\ Phys.\ J.\ C {\bf 48} (2006) 147
  doi:10.1140/epjc/s10052-006-0014-8
  [hep-ph/0512077];
A.~G.~Akeroyd, M.~A.~Diaz, M.~A.~Rivera and D.~Romero,
  Phys.\ Rev.\ D {\bf 83} (2011) 095003
  doi:10.1103/PhysRevD.83.095003
  [arXiv:1010.1160 [hep-ph]];
A.~Delgado, M.~Garcia-Pepin, M.~Quiros, J.~Santiago and R.~Vega-Morales,
  JHEP {\bf 1606} (2016) 042
  doi:10.1007/JHEP06(2016)042
  [arXiv:1603.00962 [hep-ph]].


\bibitem{exptfer}
P.~Abreu {\it et al.} [DELPHI Collaboration],
  Phys.\ Lett.\ B {\bf 507} (2001) 89
  doi:10.1016/S0370-2693(01)00449-X
  [hep-ex/0104025];
 P.~Achard {\it et al.} [L3 Collaboration],
  Phys.\ Lett.\ B {\bf 534} (2002) 28
  doi:10.1016/S0370-2693(02)01572-1
  [hep-ex/0203016];
B.~Abbott {\it et al.} [D0 Collaboration],
  Phys.\ Rev.\ Lett.\  {\bf 82} (1999) 2244
  doi:10.1103/PhysRevLett.82.2244
  [hep-ex/9811029];
S.~Chatrchyan {\it et al.} [CMS Collaboration],
  Phys.\ Lett.\ B {\bf 725} (2013) 36
  doi:10.1016/j.physletb.2013.06.043
  [arXiv:1302.1764 [hep-ex]].

\bibitem{fer1}
 H.  H. E. Haber, G. L. Kane and T. Sterling, Nucl.\ Phys.\ {B 161}, 493 (1979);
S.~Mrenna and J.~D.~Wells,
  Phys.\ Rev.\ D {\bf 63} (2001) 015006
  doi:10.1103/PhysRevD.63.015006
  [hep-ph/0001226].
E.~Gabrielli, K.~Kannike, B.~Mele, A.~Racioppi and M.~Raidal,
  Phys.\ Rev.\ D {\bf 86} (2012) 055014
  doi:10.1103/PhysRevD.86.055014
  [arXiv:1204.0080 [hep-ph]].

\bibitem{deBlas:2018tjm}
  J.~de Blas, O.~Eberhardt and C.~Krause,
  JHEP {\bf 1807} (2018) 048
  doi:10.1007/JHEP07(2018)048
  [arXiv:1803.00939 [hep-ph]].

\bibitem{HcouplingsLHC}
The ATLAS and  CMS Collaboration,
 ATLAS-CONF-2015-044,  CMS-PAS-HIG-15-002.
\bibitem{futureATLAS}
 [ATLAS Collaboration], ATL-PHYS-PUB-2013-014, "Projections for measurements of Higgs
 boson cross sections, branching ratios and coupling parameters with the ATLAS detector at a
 HL-LHC"
\bibitem{futureCMS}
[CMS Collaboration], "Projected Performance of an Upgraded CMS Detector at the LHC and
HL-LHC: Contribution to the Snowmass Process," arXiv:1307.7135.
\bibitem{Cepeda:2019klc}
  M.~Cepeda {\it et al.} [Physics of the HL-LHC Working Group],
  arXiv:1902.00134 [hep-ph].
\bibitem{ntools1} 
  U.~Ellwanger, J.~F.~Gunion and C.~Hugonie,
  JHEP {\bf 0502}, 066 (2005)
  doi:10.1088/1126-6708/2005/02/066
  [hep-ph/0406215];
  U.~Ellwanger and C.~Hugonie,
  Comput.\ Phys.\ Commun.\  {\bf 175}, 290 (2006)
  doi:10.1016/j.cpc.2006.04.004
  [hep-ph/0508022].
  U.~Ellwanger and C.~Hugonie,
  Comput.\ Phys.\ Commun.\  {\bf 177}, 399 (2007)
  doi:10.1016/j.cpc.2007.05.001
  [hep-ph/0612134];
  D.~Das, U.~Ellwanger and A.~M.~Teixeira,
  Comput.\ Phys.\ Commun.\  {\bf 183}, 774 (2012)
  doi:10.1016/j.cpc.2011.11.021
  [arXiv:1106.5633 [hep-ph]].
\bibitem{lux} 
  D.~S.~Akerib {\it et al.} [LUX Collaboration],
  Phys.\ Rev.\ Lett.\  {\bf 118}, no. 2, 021303 (2017)
  doi:10.1103/PhysRevLett.118.021303
  [arXiv:1608.07648 [astro-ph.CO]].
\bibitem{astrouncer} 
  C.~McCabe,
  Phys.\ Rev.\ D {\bf 82}, 023530 (2010)
  doi:10.1103/PhysRevD.82.023530
  [arXiv:1005.0579 [hep-ph]];
  M.~T.~Frandsen, F.~Kahlhoefer, C.~McCabe, S.~Sarkar and K.~Schmidt-Hoberg,
  JCAP {\bf 1201}, 024 (2012)
  doi:10.1088/1475-7516/2012/01/024
  [arXiv:1111.0292 [hep-ph]].

\bibitem{myastro} 
  D.~Das and U.~Ellwanger,
  JHEP {\bf 1009}, 085 (2010)
  doi:10.1007/JHEP09(2010)085
  [arXiv:1007.1151 [hep-ph]];
  D.~Das, A.~Goudelis and Y.~Mambrini,
  JCAP {\bf 1012}, 018 (2010)
  doi:10.1088/1475-7516/2010/12/018
  [arXiv:1007.4812 [hep-ph]].

\bibitem{nmssmgravitino} 
  J.~Hasenkamp and M.~W.~Winkler,
  Nucl.\ Phys.\ B {\bf 877}, 419 (2013)
  doi:10.1016/j.nuclphysb.2013.10.017
  [arXiv:1308.2678 [hep-ph]].

\bibitem{lsusy1} 
  R.~Barbieri, L.~J.~Hall, Y.~Nomura and V.~S.~Rychkov,
  Phys.\ Rev.\ D {\bf 75}, 035007 (2007)
  doi:10.1103/PhysRevD.75.035007
  [hep-ph/0607332].

\bibitem{lsusy2} 
  L.~J.~Hall, D.~Pinner and J.~T.~Ruderman,
  JHEP {\bf 1204}, 131 (2012)
  doi:10.1007/JHEP04(2012)131
  [arXiv:1112.2703 [hep-ph]].


\bibitem{lsusy3} 
  M.~Perelstein and B.~Shakya,
  Phys.\ Rev.\ D {\bf 88}, no. 7, 075003 (2013)
  doi:10.1103/PhysRevD.88.075003
  [arXiv:1208.0833 [hep-ph]].

\bibitem{lsusy4} 
  M.~Farina, M.~Perelstein and B.~Shakya,
  JHEP {\bf 1404}, 108 (2014)
  doi:10.1007/JHEP04(2014)108
  [arXiv:1310.0459 [hep-ph]].

\bibitem{loopcorrection}
 G.~Degrassi, S.~Heinemeyer, W.~Hollik, P.~Slavich and G.~Weiglein,
  Eur.\ Phys.\ J.\ C {\bf 28}, 133 (2003),
  [hep-ph/0212020];
 B.~C.~Allanach, A.~Djouadi, J.~L.~Kneur, W.~Porod and P.~Slavich,
  JHEP {\bf 0409}, 044 (2004), 
  [hep-ph/0406166];
S.~P.~Martin,
  Phys.\ Rev.\ D {\bf 75}, 055005 (2007), 
  [hep-ph/0701051];
 R.~V.~Harlander, P.~Kant, L.~Mihaila and M.~Steinhauser,
  Phys.\ Rev.\ Lett.\  {\bf 100}, 191602 (2008), 
  [Phys.\ Rev.\ Lett.\  {\bf 101}, 039901 (2008)], 
  [arXiv:0803.0672 [hep-ph]];
 S.~Heinemeyer, O.~Stal and G.~Weiglein,
  Phys.\ Lett.\ B {\bf 710}, 201 (2012), 
  [arXiv:1112.3026 [hep-ph]];
 A.~Arbey, M.~Battaglia, A.~Djouadi and F.~Mahmoudi,
  JHEP {\bf 1209}, 107 (2012), 
  [arXiv:1207.1348 [hep-ph]];
  M.~Chakraborti, U.~Chattopadhyay and R.~M.~Godbole,
  Phys.\ Rev.\ D {\bf 87}, no. 3, 035022 (2013), 
  [arXiv:1211.1549 [hep-ph]].
\bibitem{xsection}
https://twiki.cern.ch/twiki/bin/view/LHCPhysics/CERNYellowReportPageeBSMAt13TeV

\bibitem{higgsbound} 
  P.~Bechtle, O.~Brein, S.~Heinemeyer, G.~Weiglein and K.~E.~Williams,
  Comput.\ Phys.\ Commun.\  {\bf 181}, 138 (2010)
  doi:10.1016/j.cpc.2009.09.003
  [arXiv:0811.4169 [hep-ph]], 
  P.~Bechtle, O.~Brein, S.~Heinemeyer, G.~Weiglein and K.~E.~Williams,
  Comput.\ Phys.\ Commun.\  {\bf 182}, 2605 (2011)
  doi:10.1016/j.cpc.2011.07.015
  [arXiv:1102.1898 [hep-ph]],
  P.~Bechtle, O.~Brein, S.~Heinemeyer, O.~Stål, T.~Stefaniak, G.~Weiglein and K.~E.~Williams,
  Eur.\ Phys.\ J.\ C {\bf 74}, no. 3, 2693 (2014)
  doi:10.1140/epjc/s10052-013-2693-2
  [arXiv:1311.0055 [hep-ph]],
  P.~Bechtle, S.~Heinemeyer, O.~Stal, T.~Stefaniak and G.~Weiglein,
  Eur.\ Phys.\ J.\ C {\bf 75}, no. 9, 421 (2015)
  doi:10.1140/epjc/s10052-015-3650-z
  [arXiv:1507.06706 [hep-ph]].


\bibitem{atlas1} 
  G.~Aad {\it et al.} [ATLAS Collaboration],
  Eur.\ Phys.\ J.\ C {\bf 76}, no. 1, 45 (2016)
  doi:10.1140/epjc/s10052-015-3820-z
  [arXiv:1507.05930 [hep-ex]].
\bibitem{atlas2} 
  G.~Aad {\it et al.} [ATLAS Collaboration],
  JHEP {\bf 1601}, 032 (2016)
  doi:10.1007/JHEP01(2016)032
  [arXiv:1509.00389 [hep-ex]].

\bibitem{atlas3} 
  M.~Aaboud {\it et al.} [ATLAS Collaboration],
  Eur.\ Phys.\ J.\ C {\bf 78}, no. 1, 24 (2018)
  doi:10.1140/epjc/s10052-017-5491-4
  [arXiv:1710.01123 [hep-ex]].
\bibitem{atlas4} 
  M.~Aaboud {\it et al.} [ATLAS Collaboration],
  arXiv:1712.06386 [hep-ex].
\bibitem{cms1} 
 CMS Collaboration [CMS Collaboration], CMS-PAS-HIG-17-012.


\bibitem{CMS_bbH_tautau} 
  CMS Collaboration [CMS Collaboration],
  CMS-PAS-HIG-17-020.

\bibitem{CMS_bbH_bb} 
  CMS Collaboration [CMS Collaboration],
  CMS-PAS-HIG-16-018.

\bibitem{sushi}
R.~V.~Harlander, S.~Liebler and H.~Mantler,
  Comput.\ Phys.\ Commun.\  {\bf 184} (2013) 1605
  doi:10.1016/j.cpc.2013.02.006
  [arXiv:1212.3249 [hep-ph]];
R.~V.~Harlander, S.~Liebler and H.~Mantler,
  Comput.\ Phys.\ Commun.\  {\bf 212} (2017) 239
  doi:10.1016/j.cpc.2016.10.015
  [arXiv:1605.03190 [hep-ph]].
\bibitem{atlashiggspairprod}
  G.~Aad {\it et al.} [ATLAS Collaboration],
  Phys.\ Rev.\ Lett.\  {\bf 114}, no. 8, 081802 (2015)
  doi:10.1103/PhysRevLett.114.081802
  [arXiv:1406.5053 [hep-ex]];
  G.~Aad {\it et al.} [ATLAS Collaboration],
  Eur.\ Phys.\ J.\ C {\bf 75}, no. 9, 412 (2015)
  doi:10.1140/epjc/s10052-015-3628-x
  [arXiv:1506.00285 [hep-ex]];
  G.~Aad {\it et al.} [ATLAS Collaboration],
  Phys.\ Rev.\ D {\bf 92}, 092004 (2015)
  doi:10.1103/PhysRevD.92.092004
  [arXiv:1509.04670 [hep-ex]];
  The ATLAS collaboration,
  ATLAS-CONF-2016-004.
\bibitem{cmshiggspairprod}
  V.~Khachatryan {\it et al.} [CMS Collaboration],
  Phys.\ Rev.\ D {\bf 94}, no. 5, 052012 (2016)
  doi:10.1103/PhysRevD.94.052012
  [arXiv:1603.06896 [hep-ex]];
  CMS Collaboration [CMS Collaboration],
  CMS-PAS-B2G-17-006;
  CMS Collaboration [CMS Collaboration],
  CMS-PAS-HIG-17-008;
  CMS Collaboration [CMS Collaboration],
  CMS-PAS-HIG-17-009;
  A.~M.~Sirunyan {\it et al.} [CMS Collaboration],
  Phys.\ Lett.\ B {\bf 778}, 101 (2018)
  doi:10.1016/j.physletb.2018.01.001
  [arXiv:1707.02909 [hep-ex]];
  A.~M.~Sirunyan {\it et al.} [CMS Collaboration],
  JHEP {\bf 1801}, 054 (2018)
  doi:10.1007/JHEP01(2018)054
  [arXiv:1708.04188 [hep-ex]];
  A.~M.~Sirunyan {\it et al.} [CMS Collaboration],
  Phys.\ Lett.\ B {\bf 781}, 244 (2018)
  doi:10.1016/j.physletb.2018.03.084
  [arXiv:1710.04960 [hep-ex]].
\bibitem{cmscurrent}
  https://twiki.cern.ch/twiki/bin/view/CMSPublic/SummaryResultsHIG
\bibitem{Alwall:2011uj} 
  J.~Alwall, M.~Herquet, F.~Maltoni, O.~Mattelaer and T.~Stelzer,
  JHEP {\bf 1106}, 128 (2011)
  doi:10.1007/JHEP06(2011)128
  [arXiv:1106.0522 [hep-ph]].

  \bibitem{Alwall:2014hca} 
  J.~Alwall {\it et al.},
  JHEP {\bf 1407}, 079 (2014)
  doi:10.1007/JHEP07(2014)079
  [arXiv:1405.0301 [hep-ph]].

\bibitem{Ball:2012cx} 
  R.~D.~Ball {\it et al.},
  Nucl.\ Phys.\ B {\bf 867}, 244 (2013)
  doi:10.1016/j.nuclphysb.2012.10.003
  [arXiv:1207.1303 [hep-ph]].
\bibitem{Ball:2014uwa} 
  R.~D.~Ball {\it et al.} [NNPDF Collaboration],
  JHEP {\bf 1504}, 040 (2015)
  doi:10.1007/JHEP04(2015)040
  [arXiv:1410.8849 [hep-ph]].

  \bibitem{Sjostrand:2006za} 
  T.~Sjostrand, S.~Mrenna and P.~Z.~Skands,
  JHEP {\bf 0605}, 026 (2006)
  doi:10.1088/1126-6708/2006/05/026
  [hep-ph/0603175].
\bibitem{deFavereau:2013fsa} 
  J.~de Favereau {\it et al.} [DELPHES 3 Collaboration],
  JHEP {\bf 1402}, 057 (2014)
  doi:10.1007/JHEP02(2014)057
  [arXiv:1307.6346 [hep-ex]].
\bibitem{Selvaggi:2014mya} 
  M.~Selvaggi,
  J.\ Phys.\ Conf.\ Ser.\  {\bf 523}, 012033 (2014).
  doi:10.1088/1742-6596/523/1/012033
\bibitem{Mertens:2015kba} 
  A.~Mertens,
  J.\ Phys.\ Conf.\ Ser.\  {\bf 608}, no. 1, 012045 (2015).
  doi:10.1088/1742-6596/608/1/012045

\bibitem{Cacciari:2008gp} 
  M.~Cacciari, G.~P.~Salam and G.~Soyez,
  JHEP {\bf 0804}, 063 (2008)
  doi:10.1088/1126-6708/2008/04/063
  [arXiv:0802.1189 [hep-ph]].
\bibitem{Cacciari:2005hq} 
  M.~Cacciari and G.~P.~Salam,
  Phys.\ Lett.\ B {\bf 641}, 57 (2006)
  doi:10.1016/j.physletb.2006.08.037
  [hep-ph/0512210].
  \bibitem{Cacciari:2011ma} 
  M.~Cacciari, G.~P.~Salam and G.~Soyez,
  Eur.\ Phys.\ J.\ C {\bf 72}, 1896 (2012)
  doi:10.1140/epjc/s10052-012-1896-2
  [arXiv:1111.6097 [hep-ph]].
\bibitem{Maltoni:2013sma} 
  F.~Maltoni, K.~Mawatari and M.~Zaro,
  Eur.\ Phys.\ J.\ C {\bf 74}, no. 1, 2710 (2014)
  doi:10.1140/epjc/s10052-013-2710-5
  [arXiv:1311.1829 [hep-ph]].
\bibitem{Hussein:2017pjz} 
  M.~Y.~Hussein,
  arXiv:1703.03952 [hep-ph].
\bibitem{Sirunyan:2017zjc} 
  A.~M.~Sirunyan {\it et al.} [CMS Collaboration],
  Eur.\ Phys.\ J.\ C {\bf 78}, 165 (2018)
  Erratum: [Eur.\ Phys.\ J.\ C {\bf 78}, no. 6, 515 (2018)]
  doi:10.1140/epjc/s10052-018-5567-9, 10.1140/epjc/s10052-018-5769-1
  [arXiv:1709.08601 [hep-ex]].
  \bibitem{Aaboud:2017rwm} 
  M.~Aaboud {\it et al.} [ATLAS Collaboration],
  Phys.\ Rev.\ D {\bf 97}, no. 3, 032005 (2018)
  doi:10.1103/PhysRevD.97.032005
  [arXiv:1709.07703 [hep-ex]].
\bibitem{Cascioli:2014yka} 
  F.~Cascioli {\it et al.},
  Phys.\ Lett.\ B {\bf 735}, 311 (2014)
  doi:10.1016/j.physletb.2014.06.056
  [arXiv:1405.2219 [hep-ph]].
\bibitem{Campanario:2014ioa}
  F.~Campanario, M.~Kerner, L.~D.~Ninh and D.~Zeppenfeld,
  JHEP {\bf 1407} (2014) 148
  doi:10.1007/JHEP07(2014)148
  [arXiv:1405.3972 [hep-ph]].
  
\bibitem{mcfm}
https://mcfm.fnal.gov/


\end{thebibliography}
\end{document}